\documentclass[conference,compsoc]{IEEEtran}

\usepackage[numbers,sort&compress]{natbib}

\ifCLASSOPTIONcompsoc
  \usepackage[nocompress]{cite}
\else
  \usepackage{cite}
\fi
\ifCLASSINFOpdf
\else
\fi

\usepackage{stfloats}
\usepackage{amsmath}
\usepackage{amssymb}
\DeclareMathOperator*{\argmin}{arg\,min}
\usepackage{booktabs}
\usepackage{pgfplots}
\usepgfplotslibrary{groupplots}
\pgfplotsset{compat=1.18}
\definecolor{seriesblue}{HTML}{2a78d6}
\definecolor{seriesorange}{HTML}{eb6834}
\definecolor{seriesaqua}{HTML}{1baf7a}
\definecolor{seriesred}{HTML}{d6435a}
\definecolor{seriespink}{HTML}{d67ab0}
\usepackage[colorlinks=true,linkcolor=blue!60!black,citecolor=green!40!black,urlcolor=blue!60!black]{hyperref}
\usepackage{algorithm}
\usepackage{algpseudocode}
\usepackage{amsthm}
\newtheorem{definition}{Definition}
\newtheorem{lemma}{Lemma}
\newtheorem{theorem}{Theorem}
\newtheorem{corollary}{Corollary}

\begin{document}
%
\title{STEALTH: Steganographic Text Encoding Agnostic to\\ Low-Level Tensor Hardware}

\author{\IEEEauthorblockN{William Fleshman}
\IEEEauthorblockA{Army Cyber Institute\\
United States Military Academy\\
West Point, New York 10996\\
william.fleshman@westpoint.edu}
}


\maketitle

\begin{abstract}
Provably secure linguistic steganography hides a message inside text sampled from a language model, so that without the key, a stegotext is computationally indistinguishable from an honest model sample. Nearly every such scheme assumes the receiver can recompute the sender's token probabilities exactly bit for bit. We show this assumption is fragile: across three model families, bit-exact schemes decode almost none of our messages once the receiver's floating-point precision, quantization, or GPU architecture differs from the sender's, or even when the receiver merely scores a text in parallel versus one token at a time, with the same model, precision, and hardware held fixed. We present {\scshape Stealth}, a scheme built for exactly this setting. The sender draws each token with an exponential race driven by keyed pseudorandom noise, so a receiver's probability mismatch flips a token only when the race is close; a rateless spinal code, decoded by beam search, absorbs the resulting flips without the sender and receiver ever agreeing on a single distribution. We prove stegotexts remain indistinguishable from honest samples under this construction. Empirically, {\scshape Stealth} decodes every message we sent for every receiver in the sender's model family, at every precision and quantization level, on smaller models, and on a different GPU architecture. {\scshape Stealth} is even robust across different model families with separate vocabularies and tokenizers, decoding every message we tested. Our approach carries far more payload per token than prior robust schemes, and trained steganalyzers and a targeted statistical test both fail to distinguish its stegotexts from honest model output.
\end{abstract}


%
\IEEEpeerreviewmaketitle

\newcommand{\stealth}{{\upshape\scshape Stealth}}

\section{Introduction}
\label{sec:intro}

State and corporate surveillance of digital communication is pervasive~\citep{rfc7258}, and censorship regimes increasingly flag or block traffic that looks encrypted or otherwise anomalous~\citep{wu2023gfw}. While encrypted communications maintain privacy, they are easily identified. A useful alternative is covert channels: hiding messages inside communication that looks benign, so there is nothing anomalous for a censor or surveillant to flag. Steganography is the general technique, embedding secret messages into ordinary media such as images or text so the resulting cover object appears unremarkable to anyone but the intended recipient (\autoref{sec:background}). Ideally, a steganographic technique should be \textit{provably secure}, indistinguishable from the cover channel even to a warden who can run the embedding algorithm itself, rather than merely undetected by today's tools~\citep{hopper2002provably}.

Achieving this for images can be difficult since the true statistics of natural images are hard to model exactly~\citep{ker2013moving}. Modern language models (LMs) make text more tractable because a good LM provides an explicit probability distribution over realistic text continuations. Linguistic steganography uses this distribution to hide messages: the sender uses the message bits to choose each sampled token, and the receiver, recomputing the same distribution, reads the bits back from the tokens that are observed. Several such schemes are provably secure~\citep{kaptchuk2021meteor,dewitt2023imec,ding2023discop}, but only if the receiver reproduces the sender's distribution bit for bit, since the message is encoded in exactly where each sampled token falls within it (\autoref{sec:background}). This bit-exact requirement is harder to meet than it sounds: floating-point reductions are not associative, so different tensor hardware, kernels, precision, or quantization yield different results for the same model and text, and even identical hardware can disagree between generating text token by token and scoring it in a single pass~\citep{atil2024nondeterminism,yuan2025fp32,thinkingmachines2025}. In practice, sender and receiver probabilities almost always differ, and because decoding replays the sender's computation step by step, a single token whose probability differs in the last few bits can send every subsequent step down the wrong path. A recent survey of linguistic steganography lists this computational indeterminism as an open challenge~\citep{yan2026survey}.

\begin{figure*}[t]
\centering
\includegraphics[width=0.83\textwidth]{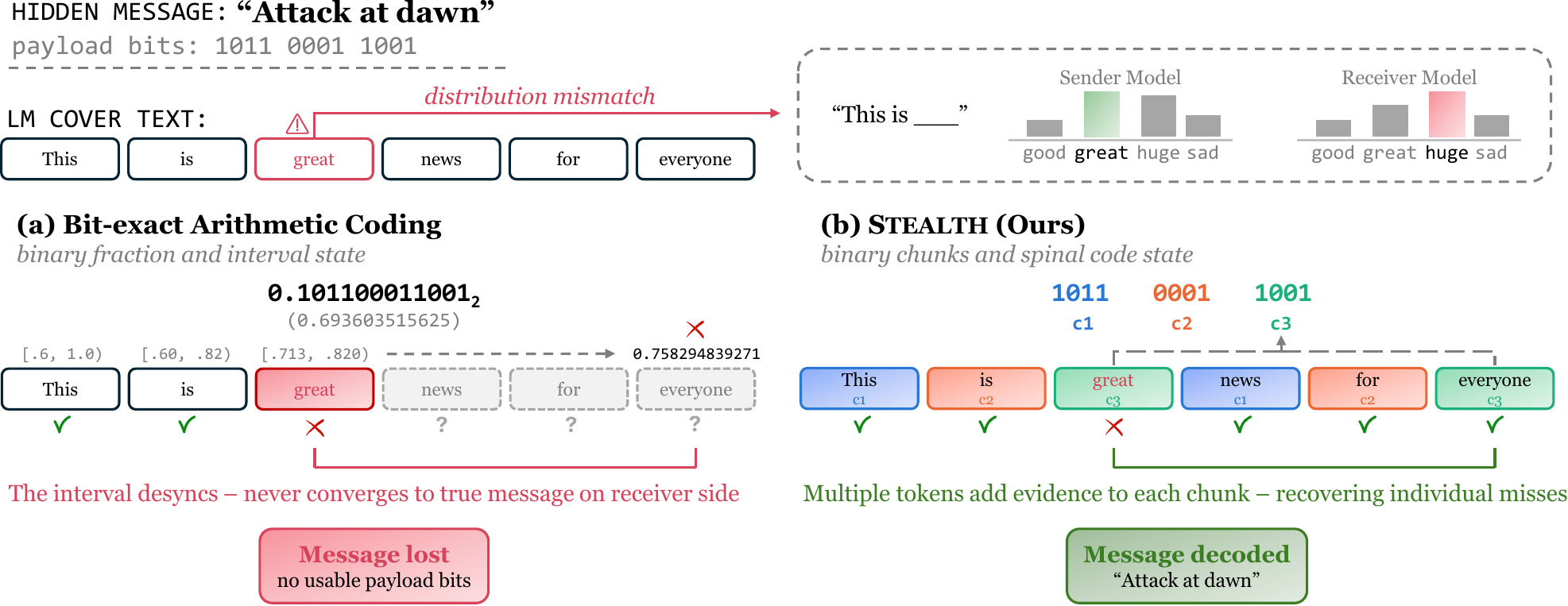}
\caption{Differences in sender and receiver token distributions cause the receiver to lose sync at a single token, desyncing bit-exact decoding schemes and losing the message entirely; \stealth{}'s spinal-code state lets later tokens supply evidence that recovers the payload despite the mismatch.}
\label{fig:overview}
\end{figure*}

We design \stealth{} to tolerate this mismatch while remaining provably secure (\autoref{fig:overview}). The sender picks each token by an \textit{exponential race} over the token distribution $p$, choosing the token that minimizes $E_i/p_i$, where $E_i$ is keyed pseudorandom exponential noise. With truly random noise, this samples exactly from $p$; with keyed noise, it is indistinguishable from doing so~\citep{christ2024undetectable,kuditipudi2024robust}. If the receiver's distribution $p'$ differs from $p$, the same noise picks a different token with probability bounded by a simple function of the total variation distance between the two distributions. The receiver scores the cover text one position at a time using the true preceding tokens as context (teacher forcing), so a flip affects only its own position and never the context of later tokens. We design the noise at each position to depend on the message through a \textit{spinal code}~\citep{perry2012spinal}, a hash chain over message chunks that the receiver decodes by an efficient beam search with a soft likelihood treating a narrowly lost race as a probable flip rather than a contradiction. Because the code is rateless, every additional token the model writes contributes more evidence toward the payload, so the sender never needs to truncate the text to fit a fixed message length. Positions in the text are measured in byte offsets, and the noise at each position can be keyed by the token's bytes rather than its vocabulary index, so nothing the receiver computes depends on the sender's tokenizer. This lets the receiver decode with an LM from an entirely different model family, with its own vocabulary and tokenizer.

Like any steganographic technique, \stealth{} is dual-use: the same mechanism that lets a dissident evade censorship can let an adversary run a covert channel undetected. Because \stealth{} removes the deployment-matching requirement that made earlier schemes fragile, it also lowers the bar for using steganography as a reliable, hard-to-detect channel between two parties with varying infrastructure. Of note, any defense that relies on sender and receiver running mismatched deployments to disrupt steganographic coordination will not hold against a scheme like \stealth{}. A sender on a datacenter GPU and a receiver on a consumer GPU running a 4-bit quantized model of the same family can still communicate, with messages provably indistinguishable from honest model output. In this paper, we:

\begin{itemize}
    \item Demonstrate that schemes relying on bit-exact computation fail under ordinary mismatches in hardware, precision, or sequence processing procedures;
    \item Propose \stealth, which tolerates these mismatches, even enabling senders with speculative decoding, or receivers with different models altogether;
    \item Prove \stealth{} is secure and produces text indistinguishable from benign model output; and
    \item Evaluate \stealth{} across three model families, five receiver configurations per family, smaller same-family receivers, receivers of other model families, and two GPU architectures, against bit-exact and robust baselines, with statistical detection tests.
\end{itemize}

\section{Background and Related Work}
\label{sec:background}

We first place \stealth{} within the steganography and covert-channel literature, then define sampling from a language model and what it means for a stegosystem built on one to be provably secure. We then explain why a receiver cannot count on recomputing the sender's distribution exactly, and cover the sampler \stealth{} uses to tolerate this, the alternatives other schemes take, the coding scheme that spreads a message across tokens, and the steganalysis literature that complements provable security.

\textbf{Steganography and covert channels.} Censors need not break encryption to block it: deep packet inspection can flag a TLS handshake or tunnel without decrypting a single byte~\citep{xue2024fingerprinting}, so censorship-resistant communication increasingly turns to covert channels that hide messages inside traffic that looks benign~\citep{dixon2016obfuscation}. These channels face their own challenge, since merely looking benign under casual inspection does not rule out being statistically distinguishable from genuinely benign traffic under closer scrutiny~\citep{cachin1998informationtheoretic}. For example, in image steganography a message can be converted to a bitstream and each bit used to replace the least significant bit of a pixel's color value~\citep{johnson1998exploring}. While imperceptible to a human eye, it is still  statistically detectable, since altering the least significant bits distorts an image's natural statistics in ways statistical steganalysis can detect, motivating decades of research into more covert embedding methods~\citep{fridrich2001reliable}. This also motivates provable security~\citep{hopper2002provably}, which we formalize in \autoref{sec:security}.

\textbf{Provably secure linguistic steganography.} An LM maps a context $x$ (the prompt and the text so far) to a distribution $p(x)$ over a vocabulary of tokens $V$ that includes an end-of-text outcome \textsc{end}; text is produced by sampling a token from $p(x)$, appending it to $x$, and repeating until \textsc{end} is sampled or a length cap is hit. Following \citet{hopper2002provably}, a stegosystem is secure if no efficient warden who chooses the hidden messages can distinguish stegotexts from samples of the cover channel; for an LM, the channel is this sampling process. LM schemes have sender and receiver independently recompute $p(x)$ and encode the message in exactly where the sampled token falls within it. Arithmetic coding embeds the message as a point in the token's cumulative interval~\citep{ziegler2019neural}, but reuses randomness across tokens and is not provably secure~\citep{kaptchuk2021meteor}. Three schemes reach this bar: Meteor masks the message with a keyed pseudorandom generator and uses the result as the sampling randomness~\citep{kaptchuk2021meteor}, iMEC embeds through a minimum-entropy coupling between the message and $p(x)$~\citep{dewitt2023imec}, and Discop partitions $p(x)$ into rotated ``copies'' and encodes the message in which copy is sampled from~\citep{ding2023discop}. All four require the receiver to reconstruct the sender's $p(x)$ bit for bit to invert the embedding.

\textbf{Why probabilities differ.} In practice, this bit-exact requirement is difficult to meet. Matrix multiplications and softmax normalizations sum many floating-point terms, and the result depends on the order they are summed in; GPUs of different architectures, different kernels (e.g.\ attention implementations chosen by sequence length), and different batch sizes sum in different orders, giving different results for the same model and input~\citep{atil2024nondeterminism,yuan2025fp32,thinkingmachines2025}. This variation is largest for mid-range probabilities~\citep{tokenprob2026}, which arise at exactly the high-entropy steps where a stegosystem embeds most of its message. Lower-precision deployments (bf16, fp16) and weight quantization (8-bit, 4-bit NF4~\citep{dettmers2023qlora}) add larger, systematic differences on top of this. Generation with a key-value cache (one forward pass per token) and teacher-forced scoring (one forward pass over the whole text) compute the same function along different paths, and can disagree even on identical hardware running identical weights~\citep{thinkingmachines2025}. \autoref{sec:exact} measures how badly these break schemes relying on bit-exact probability distributions.

\textbf{Robust steganography.} Other schemes respond to probability mismatch by giving up on recomputing $p(x)$ exactly. One approach is to never let the receiver run the model at all, embedding the message in side information decoded without access to $p(x)$~\citep{zamir2024excuse,bai2024asymmetric,guo2026feedback}. This is robust by construction, but at a much lower payload rate than a scheme that samples from the model's own distribution. Another method hardens the exact schemes, like arithmetic coding, by running sender and receiver in float64, which removes most numerical mismatch between two instances of the same model, at substantial compute cost and without addressing precision, quantization, model changes, or detectability~\citep{bauer2024deaddrop}. We compare \stealth{} against both approaches in \autoref{sec:eval}.

\textbf{The exponential race.} \stealth{} tolerates the mismatch starting with a different sampler, one whose errors stay local rather than cascading. Let $E_i \sim \mathrm{Exp}(1)$ be drawn independently for each $i \in V$. Then $\argmin_i E_i/p_i$ equals $i$ with probability exactly $p_i$. Taking logarithms, the winner minimizes $\ln E_i - \ln p_i$, the Gumbel-max trick in another form~\citep{yellott1977relationship,maddison2014astar}. Unlike inverse-CDF sampling, where a probability difference at one token can shift the intervals used to sample every other token, a mismatch here changes at most the one token whose race it affects. Aaronson's LM watermark and its successors already use this sampler, but with a fixed key and no payload~\citep{aaronson2022watermark,kuditipudi2024robust}.

\textbf{Spinal codes.} What remains is spreading a message across many tokens to absorb individual differences between sender and receiver. A spinal code~\citep{perry2012spinal} splits a message into $k$-bit chunks $c_1,\dots,c_L$ and computes a \textit{spine} $s_j = H(s_{j-1}, c_j)$ with a hash function $H$. Each spine value seeds the pseudorandom noise transmitted in place of the chunk itself. Because $s_j$ depends on every chunk up to $j$, a decoder that guesses a chunk wrong will see a mismatch from that point on, which lets it test guesses in order and keep only the best $B$ partial guesses (beam search). The code is rateless: the sender can keep producing symbols for as long as it takes the receiver to succeed.

\textbf{Detecting steganography.} A separate line of work tries to catch steganography empirically rather than prove its absence. Linguistic steganalyzers train word-level neural networks~\citep{yang2019fast,niu2019rbilstmc,yang2020densely} or fine-tuned transformers~\citep{liu2019roberta} to tell stegotext from cover text by its statistics alone, without access to the embedding scheme. \citeauthor{cao2026lpv}'s missing-tail test instead targets a specific artifact of fixed-precision decoding: interval-based schemes systematically under-sample a model's most improbable tokens~\citep{cao2026lpv}. A provable-security proof rules out detection by any efficient warden, but \autoref{sec:eval-detect} checks \stealth{} against these detectors directly.

\section{Threat Model and Goals}
\label{sec:threat}

As a running example, we will consider a comment section on a public news article as our covert channel. The article serves as context for the prompt, and the sender's model writes a comment on the article to pass information. If the message needs more text, the conversation continues with further comments, each prompted by the article and everything posted before it. The receiver reads the article and the comments off the page, in the order they were posted. The same structure fits any multi-turn channel where two parties exchange LM-generated text in public view, such as a chat, a forum thread, or messages passed between agents.

Before sending any message, the sender and receiver agree on a public \textit{prompt policy}, describing how each prompt is built from the article and the texts posted so far, and on a payload length. They also share a uniformly random key $K$. The sender runs a model $M$ on its own hardware and software, while the receiver runs a model $M' \approx M$ that may differ from $M$ in precision, quantization, GPU or software stack, model size within the same family, or model family entirely, each with its own tokenizer (\autoref{sec:design}). The receiver sees only the posted strings, not the sender's probabilities, and does not know $M$'s exact configuration.

Standing between them is a passive warden, who reads every posted text and knows everything about the scheme except the key: the model and its exact configuration, the prompt policy, and how to sample honest text from $M$ itself. We allow the warden to adaptively choose the article used as context and the hidden messages being sent. The warden wins by correctly telling whether a given conversation carries a hidden message at all.

\stealth{} aims for two properties against this warden. \textit{Security} requires that the posted texts be computationally indistinguishable from honest samples of $M$, which we formalize in \autoref{sec:security}. \textit{Robustness} requires that a receiver running $M' \neq M$ recover the correct payload from as little text as possible, and almost never output a wrong one.

\section{Mismatch Breaks Bit-Exact Schemes}
\label{sec:exact}

\begin{table}[t]
\centering
\caption{Bit-exact schemes fail under mismatch. Percent of messages decoded. $^*$: results across bf16, fp16 and fp32.}
\label{tab:exact}
\footnotesize
\setlength{\tabcolsep}{4pt}
\begin{tabular}{@{}lccc@{}}
\toprule
Receiver & Gemma & Llama & Qwen \\
\midrule
\multicolumn{4}{@{}l}{\textit{Arithmetic coding, 7--9B models}} \\
Same model, replay & 100 & 100 & 100 \\
Different GPU, replay & 0 & 0 & 0 \\
Same model, teacher forced$^*$ & 0 & 0 & 0 \\
int8, NF4, smaller model & 0 & 0 & 0 \\
\midrule
\multicolumn{4}{@{}l}{\textit{Discop, 7--9B models}} \\
Same model, replay & 100 & 100 & 100 \\
Different GPU, replay & 0 & 0 & 0 \\
Same model, teacher forced$^*$ & 0.5--1.5 & 1.5 & 0.5--1 \\
int8, NF4, smaller model & 0 & 0 & 0 \\
\midrule
\multicolumn{4}{@{}l}{\textit{Arithmetic coding, w/ float64 senders, 1.5--3B models}} \\
fp64, replay & 100 & 100 & 100 \\
Different GPU, fp64 replay & 100 & 100 & 100 \\
fp64, teacher forced & 100 & 100 & 100 \\
fp32 & 0 & 0 & 0.5 \\
bf16$^*$, int8, NF4 & 0 & 0 & 0 \\
\bottomrule
\end{tabular}
\smallskip
\end{table}

To demonstrate how ordinary mismatch breaks the bit-exact schemes of \autoref{sec:background}, we test both arithmetic coding~\citep{ziegler2019neural} and Discop~\citep{ding2023discop}, and the float64 hardening proposed by \citet{bauer2024deaddrop}. Each scheme generates comments on the same set of news articles from CNN/DailyMail~\citep{see2017cnndm} used throughout our evaluation (\autoref{sec:eval}), with the same three LM senders in bf16 (Gemma-2-9B-it~\citep{gemma2}, Llama-3.1-8B-Instruct~\citep{llama3}, and Qwen2.5-7B-Instruct~\citep{qwen25}), so that every scheme faces the same mismatch as \stealth{}. A receiver can decode either by \textit{replaying} the sender's exact computation, which requires matching its weights, precision, and hardware exactly, or by \textit{teacher forcing} the posted text in one pass on the receiver's own hardware. First, we keep sender and receiver on the same hardware (best-case scenario), and vary only the processing mode, precision, quantization, or model size, dropping to the smaller sibling in each family (Gemma-2-2b-it, Llama-3.2-3B-Instruct, Qwen2.5-1.5B-Instruct)~\citep{gemma2,llama32modelcard,qwen25}. Then, we keep configurations identical, but vary the hardware between sender and receiver. The results described below are summarized in \autoref{tab:exact}.

\textbf{Arithmetic coding.} We implemented arithmetic-coding steganography~\citep{ziegler2019neural} over the full vocabulary, with tokens sorted by probability and 32-bit interval precision. With the three 7--9B senders in bf16, replaying the sender's own generation path decodes every message for all three families, as expected. Teacher forcing the same cover text decodes none. No other receiver configuration does better: decoding in fp16 or fp32, quantizing to int8 or 4-bit NF4, or substituting a smaller model from the same family fails on every message. Decoding derailed after only 5--7 correct bits with the same model, and 1--4 with other configurations.

\textbf{Discop.} For Discop~\citep{ding2023discop}, replay again decodes all messages per family. Teacher forcing decodes $\sim 1\%$, and switching precision, quantization, or model size decodes none. At every step Discop builds a Huffman tree over the whole vocabulary, so a small probability change in even a rare token can change how the tree merges, which changes every later token's path through it.

\textbf{Hardware.} Next, we pivot from best-case to a more practical scenario where the sender and receiver have different hardware. We generated each family's comments on an NVIDIA H100 and decoded them by exact replay on an NVIDIA A6000. While both methods decode all same-GPU messages this way, across the two architectures they decode none. \autoref{tab:xgpu-tv} shows why: the total variation (TV) between the two architectures' token distributions runs 0.8--1.4\% per token, on par with the TV between generating a text step-by-step and scoring it with teacher forcing on the same GPU. Not one of the positions in any message we measured was bit-identical across different architectures.

\begin{table}[t]
\centering
\caption{Total variation per token on same GPU w/ replay vs. teacher-forcing (tf) and on different (Diff) GPUs w/ replay.}
\label{tab:xgpu-tv}
\footnotesize
\setlength{\tabcolsep}{4pt}
\begin{tabular}{@{}lcc@{}}
\toprule
Receiver & Same GPU (replay vs.\ TF) & Diff GPU (replay) \\
\midrule
Gemma 9B & 0.86\% & 0.84\% \\
Gemma 2B & 1.11\% & 1.11\% \\
Llama 8B & 0.95\% & 0.94\% \\
Llama 3B & 0.94\% & 0.91\% \\
Qwen 7B & 1.24\% & 1.23\% \\
Qwen 1.5B & 1.42\% & 1.38\% \\
\bottomrule
\end{tabular}
\end{table}

\textbf{Hardening with float64.} \citeauthor{bauer2024deaddrop} propose running both sides in float64, reasoning that the added precision shrinks numerical differences below what the coder can detect, though they stop short of claiming this holds for every device combination~\citep{bauer2024deaddrop}. We found a pitfall in replicating this with Hugging Face models: several normalization and attention operations (RMSNorm in Qwen2 and Llama, Gemma 2's eager-attention softmax) silently cast to float32 regardless of the requested dtype, so float32-sized differences survive and decoding still fails after 20--24 bits. After patching these casts, float64 senders decoded all messages by teacher forcing, and float64 text generated on an H100 decodes correctly on an A6000, for every family, including inside batches with other texts. On the hardware we tested, float64 hardening removes the mismatch it targets, though as \citeauthor{bauer2024deaddrop} themselves note, nothing guarantees this holds on configurations we did not test. 

\begin{figure}[t]
\centering
\begin{tikzpicture}
\begin{axis}[
    width=0.85\columnwidth,
    height=0.6\columnwidth,
    xlabel={Comments},
    ylabel={Detector AUC},
    ymin=0.5, ymax=1.0,
    xmin=0, xmax=200,
    xtick={0,50,100,150,200},
    ytick={0.5,0.6,0.7,0.8,0.9,1.0},
    legend pos=south east,
    legend style={font=\footnotesize},
    label style={font=\small},
    tick label style={font=\small},
    axis lines*=left,
    axis line style={gray},
    grid=major,
    grid style={gray!20, line width=0.3pt},
    tick style={gray},
]
\addplot[line width=3pt, dotted, color=seriesaqua, mark=triangle*, mark size=1.4pt] coordinates {
(1,0.545) (2,0.5725) (3,0.5889) (4,0.6035) (5,0.6091) (7,0.6431)
(10,0.6502) (15,0.7049) (20,0.7125) (30,0.7675) (40,0.7972) (50,0.8238)
(70,0.8695) (100,0.9078) (150,0.9512) (200,0.9739)
};
\addlegendentry{Gemma}
\addplot[line width=3pt, dashed, color=seriesorange, mark=square*, mark size=1.2pt] coordinates {
(1,0.5435) (2,0.5493) (3,0.569) (4,0.603) (5,0.5921) (7,0.6173)
(10,0.6456) (15,0.6724) (20,0.6895) (30,0.7259) (40,0.7547) (50,0.7768)
(70,0.8235) (100,0.8633) (150,0.9106) (200,0.9421)
};
\addlegendentry{Llama}
\addplot[line width=3pt, solid, color=seriesblue, mark=*, mark size=1.2pt] coordinates {
(1,0.5465) (2,0.5728) (3,0.5817) (4,0.5897) (5,0.6191) (7,0.6316)
(10,0.6473) (15,0.6902) (20,0.7124) (30,0.7399) (40,0.779) (50,0.8065)
(70,0.8507) (100,0.8883) (150,0.9324) (200,0.9573)
};
\addlegendentry{Qwen}
\addplot[black, dashed, line width=3pt, no marks] coordinates {(90,0.5) (90,1.0)};
\addlegendentry{Reject Null}
\end{axis}
\end{tikzpicture}
\caption{A warden can distinguish float64 comments from bf16 (AUC 0.5 is chance). The vertical line marks the median, across the three families, of the number of comments at which the null hypothesis that the sender is using bf16 becomes rejectable at the 5\% significance level.}
\label{fig:precision-auc}
\end{figure}
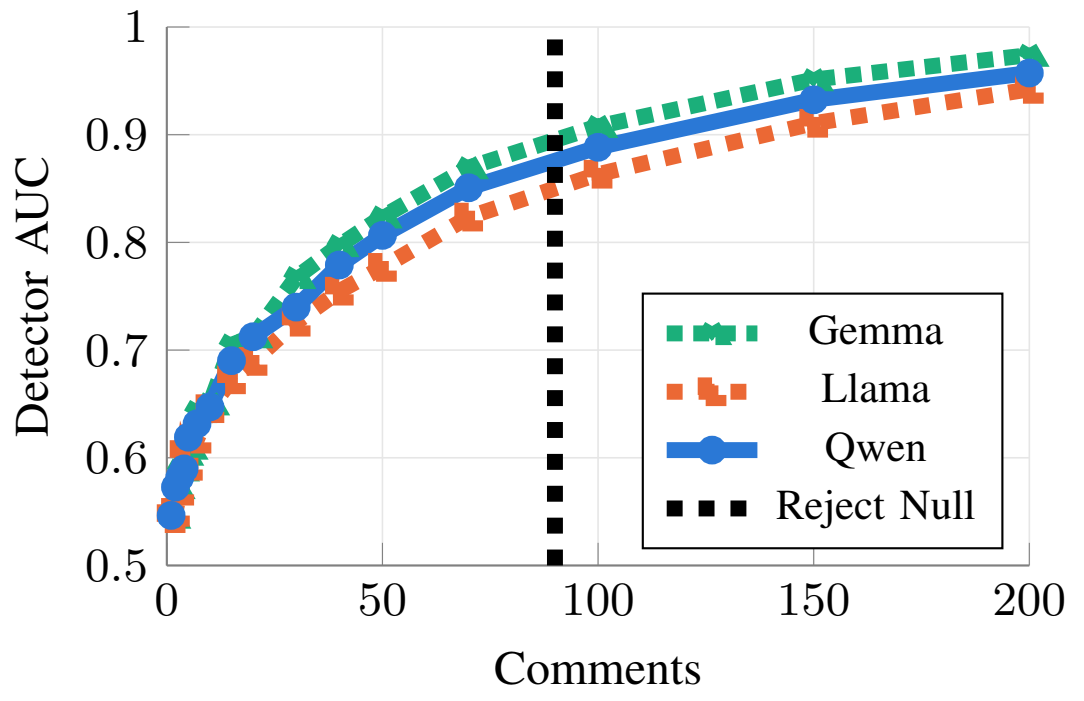

\textbf{What float64 does not fix.} The hardening keeps only one, carefully matched configuration in sync; it is not robust to anything else. A receiver at fp32 decoded at most 0.5\% of messages tested, and lower precision or quantization always failed. It is also expensive: a 7--9B model needs 61--74~GB in float64, more than a 48~GB GPU holds, so we evaluated this baseline on smaller 1.5--3B models instead. On the hardware we tested, float64 generation is 2.5--5.5$\times$ slower per token than our approach, and decoding a message takes 5.8--16.5~s against 55--71~ms for \stealth{} on the same 1.5--3B models (\autoref{sec:eval-compare}). Most importantly to our threat model, a warden that holds both a bf16 and a float64 copy of the model can also tell float64 comments from bf16 ones, by comparing how likely each comment's tokens are under the two models and accumulating that evidence across many comments (\autoref{fig:precision-auc}).

Across both coders, any mismatch collapses decoding, and the mitigation using float64 on both sides is slow, rarely deployed, and detectable given enough text. Our approach (\autoref{sec:design}) instead assumes sender and receiver will disagree, and asks only that the receiver recover the message despite that disagreement while the posted text stays indistinguishable from an honest sample.

\section{STEALTH}
\label{sec:design}

\begin{figure*}[t]
\centering
\includegraphics[width=\textwidth]{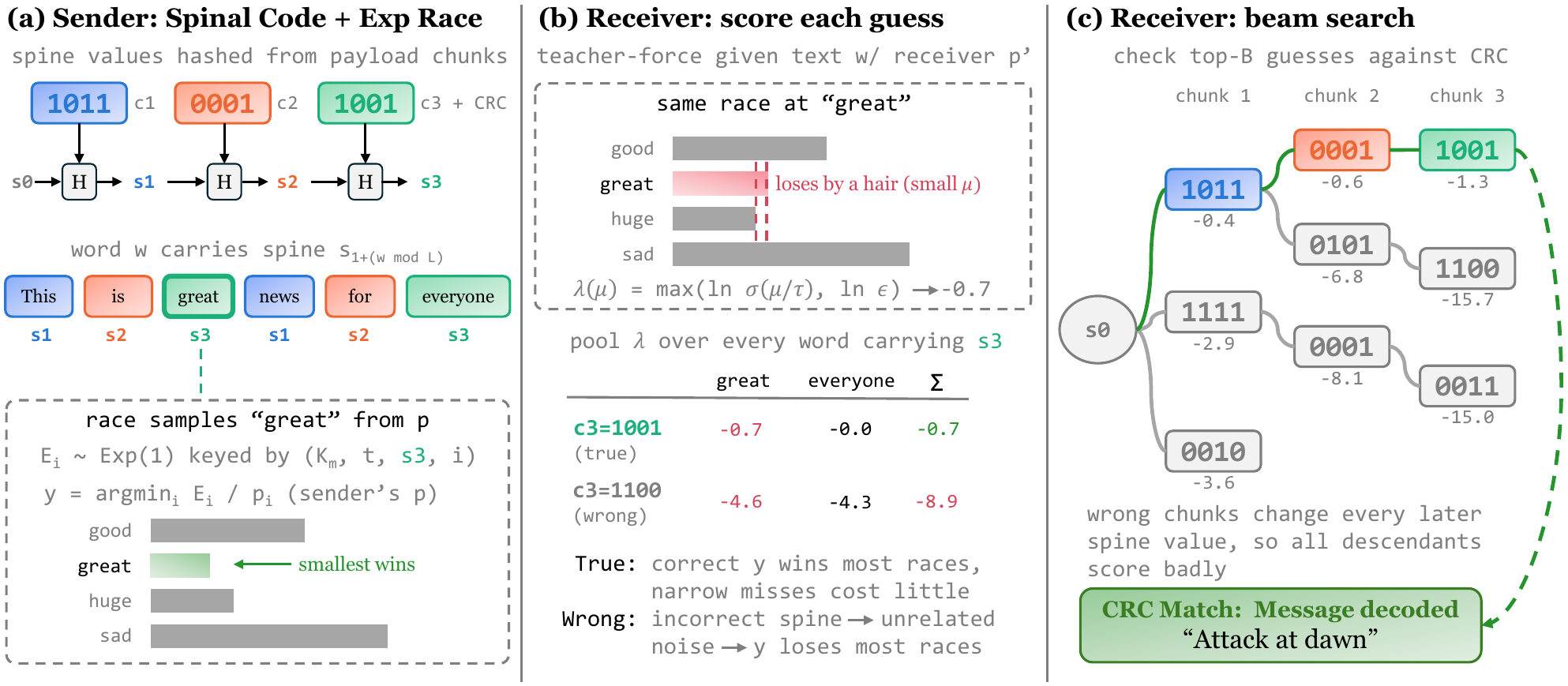}
\caption{How \stealth{} encodes and decodes the payload from \autoref{fig:overview}. (a) Payload chunks are hashed into a chain of spine values, each keying the noise for its words' tokens. (b) The receiver replays each race under its own $p'$: a narrow loss costs little, but a wrong chunk guess changes the noise and loses most races. (c) Beam search keeps the $B$ best partial guesses and outputs the best one that passes the CRC.}
\label{fig:mechanism}
\end{figure*}

We now describe \stealth{} in full. \autoref{sec:race} gives the exponential race the sender uses to sample each token, and why a mismatch at one position can flip only that position's token rather than cascading into every token that follows. \autoref{sec:spinal} shows how a spinal code spreads the payload across many such tokens, and \autoref{sec:decoding} gives the receiver's beam search over partial guesses. \autoref{sec:length} covers how a text ends and how re-tokenization is handled, and \autoref{sec:xfam} removes the sender's tokenizer from the noise entirely, letting a receiver on a different model family decode the same text. \autoref{fig:mechanism} walks through each step on the running example from \autoref{fig:overview}.

\subsection{Exponential Race}
\label{sec:race}

Each message gets its own key $K_m$, derived from the shared secret $K$ and a fresh nonce $\nu$ by a standard keyed hash function (\autoref{sec:security}); $\nu$ need not be secret, since it only has to tell the receiver which posted texts belong to this message, not protect anything by itself. At each token position, the sender draws noise $E_i \sim \mathrm{Exp}(1)$ for every $i \in V$ from a pseudorandom function keyed by $K_m$ and the token's position in the cover text the sender is writing, and outputs the token $y = \argmin_{i \in V} E_i/p_i(h)$, the exponential race introduced in \autoref{sec:background}. Repeating the race on the receiver's own distribution $p'(h)$ with the same noise reproduces $y$ whenever $p(h)$ and $p'(h)$ agree, and disagrees with probability at most $2D/(1+D)$, where $D$ is their total variation distance~\citep{daliri2024coupling}. \autoref{sec:decoding} explains how \stealth{} recovers the payload despite these disagreements. \autoref{sec:security} formalizes that the race samples exactly from $p(h)$; the disagreement bound is proved by \citeauthor{daliri2024coupling}~\citep{daliri2024coupling}.

A flip at one position never impacts the other tokens. The position $t$ is just the token's byte offset in the already-written text, so both sender and receiver read $t$ directly off the string. The noise at that position depends only on the message key, that offset, and the spine value its word carries (\autoref{sec:spinal}), never on which token was chosen there or at any earlier position, or on either party's probabilities. So, a mismatch that flips $y_t$ leaves every other position's race untouched. The bit-exact schemes of \autoref{sec:background} behave differently: their decoders carry state from one step to the next, such as an arithmetic coder's current interval, so a single wrong step corrupts every bit decoded after it. \stealth{}'s races are independent trials, each succeeding or failing on its own.

\subsection{Spinal Code}
\label{sec:spinal}

The sender frames the $n$-bit payload $m$ with a 16-bit cyclic redundancy check (CRC), which bounds how often a failed decode returns a wrong payload rather than reporting failure (\autoref{sec:eval}), and splits the result into $L$ chunks of $k=4$ bits each. A \textit{spine}~\citep{perry2012spinal} ties these chunks to the exponential race: a fixed $s_0$ starts the chain, and each later value $s_j = H(s_{j-1}, c_j)$ hashes the previous spine value together with chunk $c_j$, so $s_j$ depends on every chunk up to $j$ and changes completely if any of them does. Every token position is assigned one of the $L$ spine values when its noise is derived, so the chunk a position carries evidence for is fixed before any text is written.

Positions are assigned to spine values by counting words rather than tokens or bytes, where a word begins at the first non-whitespace byte following a whitespace byte (or at the start of the text): a token carries spine value $s_{1 + (w \bmod L)}$, where $w$ is the number of words begun before the token's first byte. The token that begins the $w$-th word (0-indexed) therefore carries $s_{1 + (w \bmod L)}$, and the rest of that word's tokens carry the next spine value; the values cycle back to $s_1$ after $s_L$. Counting words rather than tokens keeps every chunk supplied with roughly the same number of positions regardless of how the sender's tokenizer happens to split the text, and like the position itself, a word count does not depend on which tokenizer computes it, which matters once sender and receiver use different ones (\autoref{sec:xfam}).

The spinal code is rateless. Reaching $s_L$ does not end the payload's encoding, it only starts another pass over the same $L$ chunks, so a longer cover text keeps adding independent evidence for every chunk rather than running out of positions to carry them. A sender facing more mismatch, or wanting a safer margin before stopping, simply writes more text. The sender never needs to shorten, pad, or otherwise fit the payload to a fixed cover text length. This also lets the model end a text naturally, rather than cutting it off wherever the payload happens to run out, which is itself something a warden would pick up on. The full \stealth{} encoding procedure is given in Algorithm~\ref{alg:send}.

\begin{algorithm}[t]
\caption{Sender: encode payload $m$ into a cover text}
\label{alg:send}
\begin{algorithmic}[1]
\Require shared key $K$, nonce $\nu$, payload $m$, prompt, model $M$
\State $K_m \gets \mathrm{KeyedHash}(K, \nu)$
\State $c_1,\dots,c_L \gets$ chunks of $m \,\|\, \mathrm{CRC}(m)$
\State $s_0 \gets 0$; \ $s_j \gets H(s_{j-1}, c_j)$ for $j = 1,\dots,L$
\State $t, w \gets 0$; \ $\mathrm{text} \gets ()$
\Repeat
  \State $h \gets$ prompt $\|\ \mathrm{text}$
  \State $p \gets p_M(h)$
  \State $s \gets s_{1 + (w \bmod L)}$
  \State draw $E_i \sim \mathrm{Exp}(1)$ keyed by $(K_m, t, s, i)$, $i \in V$
  \State $y \gets \argmin_{i \in V} E_i / p_i$
  \State append $y$ to $\mathrm{text}$
  \State $t \gets t + \mathrm{len}_{\mathrm{bytes}}(y)$
  \State $w \gets w + \mathrm{newwords}(y)$
\Until{$y = \textsc{end}$}
\State post $\mathrm{text}$ (without \textsc{end})
\end{algorithmic}
\end{algorithm}

\subsection{Decoding}
\label{sec:decoding}

To test a guess for chunk $c_j$, the receiver teacher forces every posted text once under its own model $M'$, obtaining log-probabilities $\ell'(t)$ at each position, and recomputes the race's noise at every position assigned the guessed spine value. At such a position the observed token $y_t$ wins the race by some margin $\mu_t$, the gap between its race score and the best of the rest. A correct guess should see $y_t$ win almost every one of these positions, and lose only the narrow ones that \autoref{sec:race}'s disagreement bound predicts will flip; a wrong guess sees $y_t$ win only with probability $p'_{y_t}(h)$ at each position, and while this can be high at a single low-entropy position, it is rarely high at every position a chunk is assigned, so a wrong guess mostly loses once its evidence is pooled across all of the chunk's positions.

The receiver turns each margin into a score $\lambda(\mu_t) = \max(\ln\mathrm{sigmoid}(\mu_t/\tau),\, \ln\epsilon)$, where the temperature $\tau$ sets how much a narrow loss is discounted rather than treated as a contradiction, and the floor $\epsilon$ caps how much even a maximally wrong guess can cost at any one position. A correct guess that narrowly lost a race is charged little, while a wrong guess pays close to the floor at every position that carries real evidence. Summing $\lambda$ over the positions assigned to chunk $j$ turns this into a score for how well a guessed value of that chunk explains the text. Neither $\tau$ nor $\epsilon$ has to be fixed in advance, since the receiver already holds the whole posted text. If decoding with one setting leaves no guess passing the CRC, it can simply decode again with a different $\tau$ or $\epsilon$, for instance a larger $\tau$ if it suspects heavier mismatch than usual (\autoref{sec:eval}).

The receiver searches for the whole payload with a beam of partial guesses, extending each by every possible value of the next chunk and keeping only the best-scoring extensions. Because the spine's hash chain makes every later spine value depend on every earlier chunk, a wrong chunk corrupts the scores of everything built on top of it, and the correct sequence of chunks overtakes these corrupted guesses within the beam. Once all $L$ chunks have been guessed, the receiver outputs the highest-scoring surviving guess whose CRC passes, or reports failure if none does. \autoref{sec:spinal}'s CRC is what bounds how often this outputs a wrong payload instead of failing outright. The full decoding procedure is given in Algorithm~\ref{alg:recv}.

\begin{algorithm}[t]
\caption{Receiver: decode a payload from posted text}
\label{alg:recv}
\begin{algorithmic}[1]
\Require shared key $K$, nonce $\nu$, payload length $n$, posted text, model $M'$
\State $K_m \gets \mathrm{KeyedHash}(K, \nu)$
\State $L \gets (n+16)/4$ \Comment{CRC: 16 bits, Chunk width: 4 bits}
\State tokenize the text; append \textsc{end} if it ended naturally
\State compute $(t, w)$ for each token
\State $\ell' \gets$ log-probabilities of $M'$ over the prompt and text
\State beam $\gets \{(0, s_0, ())\}$ \Comment{score, spine value, chunks}
\For{$j = 1, \dots, L$}
  \State $P \gets$ positions with $w \bmod L = j - 1$
  \State cand $\gets \emptyset$ \Comment{candidates for this round}
  \For{$(\mathit{sc}, s, \mathit{cs}) \in$ beam, $c \in \{0,\dots,2^k{-}1\}$}
    \State $s' \gets H(s, c)$ \Comment{spine hash function}
    \State $\mathit{sc}' \gets \mathit{sc} + \sum_{t \in P} \lambda(\mu_t(K_m, t, s'))$
    \State add $(\mathit{sc}', s', \mathit{cs} \,\|\, c)$ to cand
  \EndFor
  \State beam $\gets$ the highest-scoring entries of cand
\EndFor
\State \Return the highest-scoring $\mathit{cs}$ in beam whose CRC checks, else \textsc{fail}
\end{algorithmic}
\end{algorithm}

\subsection{Additional Details}
\label{sec:length}

\textbf{Natural endings.} The \textsc{end} token takes part in the exponential race alongside every other vocabulary entry, so a text ends exactly where an honest sample from $M$ would end it, rather than wherever the payload happens to run out. While writing, the sender checks whether the text so far is already decodable by running the receiver's own algorithm on its own model. Once the text ends, the sender stops there if that check passes. Tokens written after the text first becomes decodable are not wasted, since the rateless code (\autoref{sec:spinal}) turns them into further passes over the same $L$ chunks, giving a safety margin for receivers with more mismatch than the sender assumed.

\textbf{Multiple texts.} If a model ends the text before the check passes, the message continues in another text, built from a prompt that includes every text posted so far. Position $t$ and word count $w$ keep advancing across the whole stream rather than resetting at each text's start, so a chunk's positions can span several texts exactly as they would span one.

 \textbf{Re-tokenization.} The receiver works from the posted text, not the sender's tokens directly, and re-tokenizing that text does not always reproduce the sender's token boundaries. The sender samples one token at a time, so it may write ``\texttt{\textvisiblespace hel}'' followed by ``\texttt{lo}'', which the receiver's tokenizer reads back as the single token ``\texttt{\textvisiblespace hello}''. Because position is a byte offset rather than a token index (\autoref{sec:race}), the damage stays local. The merged token sits at the first part's offset and is scored as an ordinary flip; the second part's offset is never visited, an \emph{erasure} that contributes no evidence rather than wrong evidence. Every later token keeps the sender's offsets, so the rest of the stream decodes normally, and the rateless code's redundancy (\autoref{sec:spinal}) absorbs the flip and erasure like any other.

\subsection{Decoding Across Model Families}
\label{sec:xfam}

We alter our encoding and decoding procedures slightly for scenarios where the sender and receiver are using different model families. Position $t$ and word count $w$ are defined on the posted text itself, so nothing about them ties the receiver to the sender's tokenizer, and \autoref{sec:length} already covers a receiver that re-tokenizes the same text differently. What still does tie the two together is the noise's fourth input, the vocabulary index $i$ in $(K_m, t, s, i)$. Index $i$ names a token in the sender's vocabulary, and a model of another family has its own, unrelated vocabulary, with its own index for the same string. A receiver on another model family has no way to know which of its indices corresponds to which of the sender's.

The fix is to key the noise by the token's own bytes instead of its index. Writing $\kappa(i)$ for a fixed hash of token $i$'s byte string, the sender and receiver both draw noise from $(K_m, t, s, \kappa(i))$ rather than $(K_m, t, s, i)$; only this one input to the keyed pseudorandom function changes, in both Algorithm~\ref{alg:send} and Algorithm~\ref{alg:recv}. Because $\kappa$ depends only on a token's spelling, the sender's and receiver's vocabularies agree on $\kappa(i)$ for every string they share, even though they disagree on $i$ itself. This requires $\kappa$ to be injective on the sender's vocabulary, so that two of its tokens never draw the same noise. Some vocabularies contain several tokens with the same spelling, such as Gemma's byte-fallback tokens, which duplicate ordinary single-byte tokens. Among such a group, the ordinary token, the one any tokenizer produces for that string, is keyed by its bytes, and each of the others by a hash of its index under a separate domain tag, with \textsc{end} given its own reserved key.

The receiver tokenizes the posted text with its own tokenizer and its own vocabulary, and otherwise decodes exactly as in \autoref{sec:decoding}. Wherever the two tokenizers split the text the same way, $\kappa(i)$ lines the receiver's token up with the sender's and the race scores it normally; wherever they split it differently, the mismatch is just a re-tokenization, handled the same flip-and-erasure way as \autoref{sec:length}. 

\section{Security Analysis}
\label{sec:security}

A warden watching the conversation sees only the sender's posted texts, never whether or how a receiver decodes them. What has to be shown is that those texts are indistinguishable from an honest conversation with the sender's own model, down to where each text ends.

\subsection{Security Definition}
\label{sec:secdef}

Write $\hat p(h)$ for the distribution the sender's model actually assigns a context $h$, computed from its own float32 log-probabilities rather than some idealized real-valued one. An \emph{honest text} is one sampled from $\hat p$, token by token, until \textsc{end}; an \emph{honest conversation} chains such texts together under the public prompt policy, each one's prompt built from everything posted before it. \stealth{}'s stegotext is secure if no warden can tell it apart from an honest conversation, even when it gets to choose the payloads and context adaptively, the chosen-hiddentext notion of security introduced by \citeauthor{hopper2002provably}~\citep{hopper2002provably} and instantiated here for our setting.

\begin{definition}[Security]
\label{def:security}
A warden $A$ queries an oracle with a payload $m$, a nonce $\nu$, and a starting context $h_0$, and gets back either the sender's posted texts for $m$ (the stego oracle $O_S$), or an honest conversation under the same stopping rule ($O_H$). $A$ chooses every query adaptively, knowing the outcome of every earlier one, but never reuses a nonce. Its advantage is
\[
\mathrm{Adv}(A) = \bigl|\Pr[A^{O_S} = 1] - \Pr[A^{O_H} = 1]\bigr|.
\]
\stealth{} is secure against $A$ to the extent $\mathrm{Adv}(A)$ is small; \autoref{thm:fixed} bounds it concretely in terms of the warden's number of queries $q$ and total tokens $T$.
\end{definition}

For our experiments we use HMAC-SHA256~\citep{bellare2006hmac} for \autoref{sec:race}'s keyed hash and the ChaCha20 block function~\citep{bernstein2008chacha} for its keyed pseudorandom function (PRF). We assume both are PRFs with advantage $\epsilon_{\mathrm{HMAC}}(q)$ against $q$ queries and $\epsilon_{\mathrm{ChaCha}}(r)$ against $r$ block evaluations.

\subsection{Security Theorem}
\label{sec:secthm}

\begin{lemma}[Exponential race~\citep{yellott1977relationship,maddison2014astar}]
\label{lem:race}
For a distribution $p(h)$ and independent $E_i \sim \mathrm{Exp}(1)$ for every $i$ with $p_i(h) > 0$, the race winner $W = \argmin_i E_i/p_i(h)$ satisfies $\Pr[W = i] = p_i(h)$, and the winning score $\min_i E_i/p_i(h)$ is itself $\mathrm{Exp}(1)$, independent of $W$. Proof in Appendix~\ref{app:proofs}.
\end{lemma}

This is the exact-sampling half of the claim from \autoref{sec:race}; its other half, the bound on how often a mismatched receiver's race disagrees, is proved by \citeauthor{daliri2024coupling}~\citep{daliri2024coupling}.

Lemma~\ref{lem:race} needs independent noise at every position, but the inputs \stealth{} feeds its noise function depend on the sender's own earlier tokens. The next lemma shows that, with the keyed PRF replaced by a truly random function, this never matters: the noise read at each position is fresh given everything before it. \autoref{thm:fixed} then returns to the real scheme, paying for the PRF assumption.

\begin{lemma}[Freshness]
\label{lem:fresh}
Let $\mathcal{H}_t$ be everything the sender has computed before choosing the token at position $t$. Replacing \autoref{sec:race}'s keyed pseudorandom function by a truly random function, the noise it reads to choose the token at position $t$ is independent uniform, independent of $\mathcal{H}_t$ as well. Proof in Appendix~\ref{app:proofs}.
\end{lemma}

Nothing in the proof needs the payload to be secret, so the warden may choose it. Because $t$ is itself part of the noise's input, two positions sharing a spine value still draw independent noise from each other. The argument also survives \autoref{sec:xfam}'s byte-keyed variant unchanged, since $\kappa$ is injective on the sender's vocabulary and distinct tokens still read distinct inputs there too.

\begin{corollary}[Ideal-world texts are honest]
\label{cor:honest}
With the keyed PRF replaced by a truly random function and the race computed exactly on continuous uniforms, let the sender keep generating texts indefinitely under the public prompt policy, ignoring its own stopping rule. The resulting sequence of texts has the same law as an honest conversation. Proof in Appendix~\ref{app:proofs}.
\end{corollary}

\begin{theorem}[Security]
\label{thm:fixed}
If the number of texts per hidden message is fixed in advance, or set by any rule that does not use the key, and the honest oracle uses the same rule, then for $q$ queries, $T$ tokens in total, and at most $r$ ChaCha20 evaluations per message,
\[
\mathrm{Adv}(A) \le \epsilon_{\mathrm{HMAC}}(q) + q\,\epsilon_{\mathrm{ChaCha}}(r) + T\delta,
\]
where $\delta$ is the per-token disagreement probability between the implemented race and the ideal one (\autoref{sec:fp}). Proof in Appendix~\ref{app:proofs}.
\end{theorem}

Every text ends where the sender's model would have ended an honest sample, so the only quantity in a message that can depend on the key at all is how many texts it took. Other distribution-preserving schemes end a text where the message ends rather than where the model would. Arithmetic coding, Meteor, and Discop stop once the message's bits are exhausted, and iMEC stops once the receiver's posterior over the message is confident enough~\citep{ziegler2019neural,kaptchuk2021meteor,dewitt2023imec,ding2023discop}; none of these coincide with where an honest sample would end. Each could pad with honest, message-independent text until \textsc{end}, but that padding carries nothing; in \stealth{}, every token up to the natural ending also adds evidence toward the payload.

\autoref{thm:fixed} covers a sender that posts a fixed number of texts per query, independent of the key, and simply fails the message if that was not enough. We will see in \autoref{sec:eval-robust}, that a fixed count of two would have failed none of the 600 messages we evaluated: 33 Gemma, 3 Llama, and 1 Qwen message needed a second comment, and none needed a third. A sender that instead stops as soon as the decodability check of \autoref{sec:length} passes posts less text, but its number of texts $N$ can depend on the key. By coupling it with a sender fixed at $k$ texts, a warden gains at most $\Pr[N \ne k]$, which for $k=1$ is the 6\% of messages that needed a second comment.

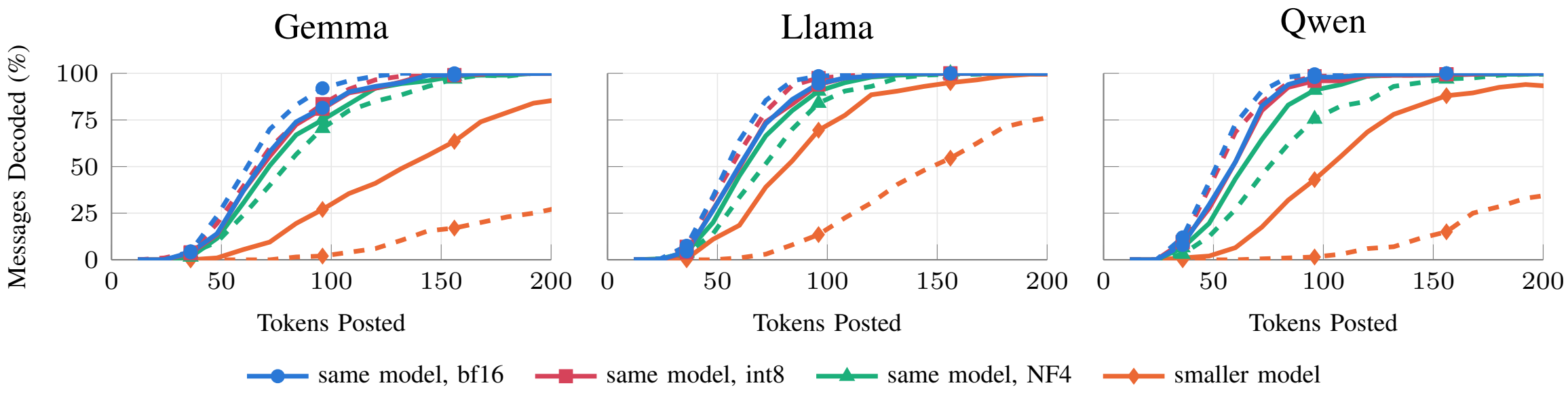
\begin{figure*}[t]
\centering
\pgfplotsset{
  every axis/.append style={
    width=0.33\textwidth, height=3.4cm,
    xlabel={Tokens Posted}, xmin=0, xmax=200,
    ymin=0, ymax=100, ytick={0,25,50,75,100},
    tick label style={font=\scriptsize}, label style={font=\scriptsize},
    axis lines*=left,
    axis line style={gray}, tick style={gray},
    grid=major, grid style={gray!20, line width=0.3pt},
  },
  sweepline/.style n args={3}{line width=1.3pt, color=#1, #2, mark=#3,
    mark size=1.3pt, mark repeat=5, mark phase=3, mark options=solid},
}
\begin{tikzpicture}
\begin{groupplot}[group style={group size=3 by 1, horizontal sep=0.55cm}]
\nextgroupplot[ylabel={Messages Decoded (\%)}, title={Gemma}]
\addplot[sweepline={seriesorange}{solid}{diamond*}]  table[x=tokens, y=smaller_t1,  col sep=comma] {figures/data/sweep_gemma.csv};
\addplot[sweepline={seriesorange}{dashed}{diamond*}] table[x=tokens, y=smaller_t01, col sep=comma] {figures/data/sweep_gemma.csv};
\addplot[sweepline={seriesaqua}{solid}{triangle*}]  table[x=tokens, y=same_nf4_t1,   col sep=comma] {figures/data/sweep_gemma.csv};
\addplot[sweepline={seriesaqua}{dashed}{triangle*}] table[x=tokens, y=same_nf4_t01,  col sep=comma] {figures/data/sweep_gemma.csv};
\addplot[sweepline={seriesred}{solid}{square*}]   table[x=tokens, y=same_int8_t1,  col sep=comma] {figures/data/sweep_gemma.csv};
\addplot[sweepline={seriesred}{dashed}{square*}]  table[x=tokens, y=same_int8_t01, col sep=comma] {figures/data/sweep_gemma.csv};
\addplot[sweepline={seriesblue}{solid}{*}]  table[x=tokens, y=same_bf16_t1,  col sep=comma] {figures/data/sweep_gemma.csv};
\addplot[sweepline={seriesblue}{dashed}{*}] table[x=tokens, y=same_bf16_t01, col sep=comma] {figures/data/sweep_gemma.csv};

\nextgroupplot[title={Llama}, yticklabels=\empty]
\addplot[sweepline={seriesorange}{solid}{diamond*}]  table[x=tokens, y=smaller_t1,  col sep=comma] {figures/data/sweep_llama.csv};
\addplot[sweepline={seriesorange}{dashed}{diamond*}] table[x=tokens, y=smaller_t01, col sep=comma] {figures/data/sweep_llama.csv};
\addplot[sweepline={seriesaqua}{solid}{triangle*}]  table[x=tokens, y=same_nf4_t1,   col sep=comma] {figures/data/sweep_llama.csv};
\addplot[sweepline={seriesaqua}{dashed}{triangle*}] table[x=tokens, y=same_nf4_t01,  col sep=comma] {figures/data/sweep_llama.csv};
\addplot[sweepline={seriesred}{solid}{square*}]   table[x=tokens, y=same_int8_t1,  col sep=comma] {figures/data/sweep_llama.csv};
\addplot[sweepline={seriesred}{dashed}{square*}]  table[x=tokens, y=same_int8_t01, col sep=comma] {figures/data/sweep_llama.csv};
\addplot[sweepline={seriesblue}{solid}{*}]  table[x=tokens, y=same_bf16_t1,  col sep=comma] {figures/data/sweep_llama.csv};
\addplot[sweepline={seriesblue}{dashed}{*}] table[x=tokens, y=same_bf16_t01, col sep=comma] {figures/data/sweep_llama.csv};

\nextgroupplot[title={Qwen}, yticklabels=\empty]
\addplot[sweepline={seriesorange}{solid}{diamond*}]  table[x=tokens, y=smaller_t1,  col sep=comma] {figures/data/sweep_qwen.csv};
\addplot[sweepline={seriesorange}{dashed}{diamond*}] table[x=tokens, y=smaller_t01, col sep=comma] {figures/data/sweep_qwen.csv};
\addplot[sweepline={seriesaqua}{solid}{triangle*}]  table[x=tokens, y=same_nf4_t1,   col sep=comma] {figures/data/sweep_qwen.csv};
\addplot[sweepline={seriesaqua}{dashed}{triangle*}] table[x=tokens, y=same_nf4_t01,  col sep=comma] {figures/data/sweep_qwen.csv};
\addplot[sweepline={seriesred}{solid}{square*}]   table[x=tokens, y=same_int8_t1,  col sep=comma] {figures/data/sweep_qwen.csv};
\addplot[sweepline={seriesred}{dashed}{square*}]  table[x=tokens, y=same_int8_t01, col sep=comma] {figures/data/sweep_qwen.csv};
\addplot[sweepline={seriesblue}{solid}{*}]  table[x=tokens, y=same_bf16_t1,  col sep=comma] {figures/data/sweep_qwen.csv};
\addplot[sweepline={seriesblue}{dashed}{*}] table[x=tokens, y=same_bf16_t01, col sep=comma] {figures/data/sweep_qwen.csv};
\end{groupplot}
\end{tikzpicture}

\vspace{2pt}
\begin{tikzpicture}
\begin{axis}[
  hide axis, width=0pt, height=0pt, scale only axis,
  xmin=0, xmax=1, ymin=0, ymax=1,
  legend columns=4, legend style={draw=none, font=\scriptsize, /tikz/every even column/.append style={column sep=6pt}},
]
\addlegendimage{sweepline={seriesblue}{solid}{*}}   \addlegendentry{same model, bf16}
\addlegendimage{sweepline={seriesred}{solid}{square*}}    \addlegendentry{same model, int8}
\addlegendimage{sweepline={seriesaqua}{solid}{triangle*}}   \addlegendentry{same model, NF4}
\addlegendimage{sweepline={seriesorange}{solid}{diamond*}} \addlegendentry{smaller model}
\end{axis}
\end{tikzpicture}

\caption{Messages decoded against tokens posted. Solid lines: $\tau=1.0$; dashed: $\tau=0.1$. Smaller models: Gemma 2B, Llama 3B, Qwen 1.5B. The axis stops at 200 tokens; \autoref{tab:sweep} gives the full counts.}
\label{fig:sweep}
\end{figure*}

\subsection{Finite-Precision Error}
\label{sec:fp}

\autoref{thm:fixed}'s remaining term, $\delta$, bounds how often \autoref{sec:race}'s implemented race disagrees with the mathematically ideal one it is supposed to compute. Write $\mathrm{Race}(\ell, W)$ for the token the sender's implementation outputs from log-probabilities $\ell$ and raw 64-bit noise words $W$, with every score computed in float64, and $\mathrm{IdealRace}(\ell, U)$ for the same race run on exactly continuous uniforms $U$, each drawn inside its word's $2^{-64}$-wide cell. The implementation departs from the ideal in two ways: rounding, since each computed score can be off by up to some $\eta$ from its exact value, and discretization, since each uniform is read from the midpoint of its $2^{-64}$-wide cell rather than sampled continuously.

\begin{lemma}[Floating-point race]
\label{lem:fp}
Let $\omega = 2^{-64}$, $W_i$, $i \in V$, be independent uniform 64-bit words, $Z_i$ independent uniforms on $[0,1)$, and $U_i = \omega(W_i+Z_i)$, which are themselves independent uniforms on $[0,1)$. Suppose the implementation reads noise $\hat E_i = -\ln(1-(W_i+\tfrac12)\omega)$ from the midpoint of $W_i$'s cell and computes every log score within $\eta$ of $\ln\hat E_i - \ell_i$, and let $\rho = e^{2\eta}-1$. Then for every $\ell$ and every $a>0$ with $\rho(a+\tfrac12) \le \tfrac14$,
\begin{multline*}
\Pr[\mathrm{Race}(\ell,W) \ne \mathrm{IdealRace}(\ell,U)] \\ \le (6+2e)\,a\,|V|\omega + 8|V|\omega + e^{-a} + \tfrac32\rho.
\end{multline*}
Proof in Appendix~\ref{app:proofs}.
\end{lemma}

Our kernel's rounding error is $\eta \approx 3\times10^{-14}$. With $|V| \le 2^{18}$, true of every vocabulary across all three model families, and $a=30$, the bound gives $\delta \le 5.2\times10^{-12}$ per token, about $1.6\times10^{-9}$ over a 300-token message. By \autoref{thm:fixed}, finite precision adds at most $T\delta$ to a warden's advantage, so even a warden that observes a million such messages gains less than $2\times10^{-3}$ from it.

\textbf{No missing tail.} \citeauthor{cao2026lpv}~\citep{cao2026lpv} detect arithmetic coding and Meteor from the probability mass their finite-precision intervals can never reach. The exponential race has no such gap: it never builds a CDF, since each token draws its own noise, and scores are compared in log space, so a token with probability $10^{-30}$ keeps its log-probability at full precision rather than being rounded to a zero-width interval. The race loses probability only where a token would need noise below the smallest representable value, $2^{-65}$, to win. On Qwen2.5 this unreachable mass totals $1.4\times10^{-17}$ per token, against $1.7\times10^{-6}$ for 32-bit arithmetic coding on the same model. This mass is part of the disagreement $\delta$ bounded by Lemma~\ref{lem:fp}, so \autoref{thm:fixed} already accounts for it. We directly run their detection test in \autoref{sec:eval-detect}.

\section{Evaluation}
\label{sec:eval}

\subsection{Setup}
\label{sec:eval-setup}

\textbf{Models.} Senders use the same 7--9B models from \autoref{sec:exact}, each in bf16, run both plainly and with speculative decoding using the smallest model of the same family as the draft (Gemma 2B, Llama 3.2 1B, Qwen 0.5B). Receivers cover the sender's own weights at bf16, fp16, fp32, int8, and 4-bit NF4, the smaller models of the same family, and for \autoref{sec:eval-xfam}, the large models of the other two families reading a byte-keyed sender. Because bf16, fp16, and fp32 gave nearly identical results throughout, the tables report bf16 only.

\textbf{Hardware.} Experiments are run on two NVIDIA architectures, Ampere (RTX A6000) and Hopper (H100). To isolate mismatch across hardware from mismatch across precision or model size, we also decode each architecture's saved texts on the other.

\textbf{Channel.} The sender's prompt is a request to comment on one of 200 CNN/DailyMail articles~\citep{see2017cnndm}, with further comments posted if the first does not carry enough evidence to decode. Texts are sampled at temperature 1 for all models.

\textbf{Code parameters.} Messages are 32-bit payloads protected by a 16-bit CRC, spun with chunk width $k=4$ (so $L=12$ chunks) and beam width $B=16$, and decoded with the decoder's score floor set to $\epsilon=10^{-3}$. We report both of the decoder's operating points, $\tau=0.1$ and $\tau=1.0$, and a cascade that falls back to $\tau=1.0$ only if $\tau=0.1$ passes no guess through the CRC. The sender's own stopping check always uses $\tau=0.1$; it runs every 12 tokens, and once it passes, the sender writes 12 more tokens as a margin before stopping at the end of the current comment. Our sweeps decode prefixes at the same 12-token granularity.

\textbf{Metrics.} \emph{Success} is whether the decoded payload matches the sent one, over the 200 messages per condition. \emph{Flip rate} is the share of tokens where the race winner under the receiver's distribution differs from the winner under the sender's, and \emph{TV} is the mean total variation distance between their per-token distributions. Both measure how far the receiver's distribution is from the sender's, which explains why more mismatch needs more tokens. \emph{Tokens to $N$\%} is the shortest prefix length at which that fraction of messages decode (\autoref{fig:sweep} plots the full curves).

\subsection{Robustness Under Mismatch}
\label{sec:eval-robust}

With \stealth{}, every receiver we tested decodes every message given enough text, across every precision, quantization level, and same-family model size; 200 of 200 bounds each configuration's failure rate below 1.9\% (95\% Wilson interval). \autoref{fig:sweep} sweeps the decoder over prefixes of long streams written and decoded on the A6000, and \autoref{tab:sweep} reports how much text each configuration needs. The worst case was with Gemma 2B, where no message needed more than 384 tokens.

\begin{table}[t]
\centering
\caption{Tokens posted to reach 90\% / 99\% / 100\% of messages decoded, using the cascade decoder.}
\label{tab:sweep}
\footnotesize
\setlength{\tabcolsep}{4pt}
\begin{tabular}{@{}lccc@{}}
\toprule
Receiver & 90\% & 99\% & 100\% \\
\midrule
Gemma 9B bf16 & 96 & 132 & 132 \\
Gemma 9B int8 & 108 & 132 & 144 \\
Gemma 9B NF4 & 120 & 156 & 192 \\
Gemma 2B & 228 & 372 & 384 \\
\midrule
Llama 8B bf16 & 84 & 108 & 132 \\
Llama 8B int8 & 84 & 120 & 132 \\
Llama 8B NF4 & 96 & 132 & 144 \\
Llama 3B & 132 & 192 & 252 \\
\midrule
Qwen 7B bf16 & 72 & 96 & 156 \\
Qwen 7B int8 & 84 & 96 & 180 \\
Qwen 7B NF4 & 96 & 132 & 204 \\
Qwen 1.5B & 180 & 300 & 300 \\
\bottomrule
\end{tabular}
\end{table}

\textbf{What drives the cost.} Within each family, the token cost in \autoref{tab:sweep} tracks flip rate (\autoref{tab:main}) rather than precision or quantization: int8 receivers, whose flips stay under 5\%, need little more than a same-configuration receiver, while NF4 and the smaller models, whose flips run from 8\% to 37\%, need more text, increasingly so as flips rise. Each flip costs evidence rather than adding it, so the spinal code's beam search needs a larger margin of correct tokens to recover.

\textbf{Decoding at the sender's stopping point.} A sender cannot see its receiver's flips; it knows only when its own model can decode. \autoref{tab:main} evaluates the simplest stopping rule built on that knowledge: the sender stops once its same-model stopping check succeeds, plus 12 more tokens, and we measure how often each receiver decodes on the sender's GPU at that point. A receiver matching the sender's configuration decodes every message in every family, since it repeats the sender's own check. Receivers close to that configuration nearly always do too: with the cascade, fp16 and fp32 lose at most one message per family, and int8 at most four. Receivers farther away do not: NF4 decodes 93.5--99.5\% of messages, and the smaller same-family models only 42--87.5\%. The sender's check alone is therefore a reliable signal only for receivers near its own configuration; to reach one farther away, the sender can post additional passes beyond the one its check calls for or, if it knows the receiver's model, run that model for its check. \begin{table}[t]
\centering
\caption{Percent of messages decoded at the sender's stopping point (its own check plus 12 tokens), with flip rate.}
\label{tab:main}
\footnotesize
\setlength{\tabcolsep}{4pt}
\begin{tabular}{@{}llcccc@{}}
\toprule
Family & Receiver & Flips (\%) & $\tau=0.1$ & $\tau=1.0$ & Cascade \\
\midrule
Gemma & 9B bf16 & 0.9 & 100 & 97.5 & 100 \\
 & 9B int8 & 3.8 & 98 & 96 & 98 \\
 & 9B NF4 & 8.4 & 87.5 & 91 & 93.5 \\
 & 2B bf16 & 31.5 & 8.5 & 42 & 42 \\
\midrule
Llama & 8B bf16 & 1.0 & 100 & 99.5 & 100 \\
 & 8B int8 & 4.9 & 100 & 99.5 & 100 \\
 & 8B NF4 & 10.2 & 93.5 & 97.5 & 98 \\
 & 3B bf16 & 23.5 & 54 & 87.5 & 87.5 \\
\midrule
Qwen & 7B bf16 & 1.2 & 100 & 100 & 100 \\
 & 7B int8 & 4.2 & 100 & 99.5 & 100 \\
 & 7B NF4 & 12.1 & 96.5 & 99.5 & 99.5 \\
 & 1.5B bf16 & 36.9 & 32.5 & 84 & 84 \\
\bottomrule
\end{tabular}
\end{table}

Neither operating point wins everywhere: $\tau=1.0$ decodes up to five fewer messages than $\tau=0.1$ on the same receiver, while $\tau=0.1$ collapses on the smaller same-family receivers, decoding 8.5--54\% of messages where $\tau=1.0$ decodes 42--87.5\%. The cascade matches or beats the better setting on every row, and beats both on Gemma and Llama NF4, as the two settings tend to fail on different messages.

\textbf{Across GPU architectures.} \autoref{tab:sweep} and \autoref{fig:sweep} decode texts on the same A6000 that wrote them. To isolate the hardware, we took the H100's own sweep texts and decoded them on both GPUs, with the cascade decoder and every receiver configuration; \autoref{tab:xhw} gives the A6000 side. Every configuration, including NF4 and the smaller same-family models, still reaches 100\% of messages decoded. Moving the receiver from the H100 to the A6000 changes no configuration's token counts by more than two 12-token steps, in either direction: the 90\% column is unchanged for 10 of 12 receivers, and 5 of 12 are unchanged in every column, so the hardware mismatch adds no systematic token cost on top of precision and quantization. The new GPU changes about 1\% of race winners at bf16 and 4--6\% at int8 relative to the same receiver on the H100, yet the flip rate against the sender stays within 0.3 points of \autoref{tab:main}'s for every receiver, and fp32 picks the same winner at every position on both architectures. Bit-exact schemes have no such slack: across the same two GPUs, arithmetic coding and Discop decode none of their messages, even by exact replay (\autoref{sec:exact}).

\begin{table}[t]
\centering
\caption{Tokens posted to reach 90\% / 99\% / 100\% of messages decoded with the cascade decoder, reading H100-written stegotext on an A6000.}
\label{tab:xhw}
\footnotesize
\setlength{\tabcolsep}{4pt}
\begin{tabular}{@{}lccc@{}}
\toprule
Receiver & 90\% & 99\% & 100\% \\
\midrule
Gemma 9B bf16 & 96 & 120 & 180 \\
Gemma 9B int8 & 96 & 144 & 180 \\
Gemma 9B NF4 & 120 & 156 & 180 \\
Gemma 2B & 228 & 300 & 324 \\
Llama 8B bf16 & 84 & 96 & 132 \\
Llama 8B int8 & 84 & 108 & 132 \\
Llama 8B NF4 & 96 & 132 & 156 \\
Llama 3B & 144 & 180 & 180 \\
Qwen 7B bf16 & 84 & 96 & 96 \\
Qwen 7B int8 & 84 & 96 & 96 \\
Qwen 7B NF4 & 96 & 108 & 168 \\
Qwen 1.5B & 168 & 252 & 264 \\
\bottomrule
\end{tabular}
\end{table}

\textbf{Efficient samplers.} Because \stealth{} tolerates mismatch, the sender is free to generate however is most efficient on its own hardware, even along computation paths the receiver never reproduces. Speculative decoding~\citep{leviathan2023speculative,chen2023speculative} is one example. Each round, a small draft model runs the same keyed race on its own distribution for a few tokens; the sender's model then scores all of them in one forward pass, runs its own race at each position with the same noise, and keeps the drafted tokens up to the first position where its race picks a different token, emitting its own winner there, as in drafter-invariant speculative decoding~\citep{daliri2024coupling}.%
\begin{table*}[t]
\centering
\caption{Cross-family decoding with byte-keyed noise. Aligned: share of the receiver's tokens matching a sender token's byte span. Flips: share of aligned tokens whose race winner differs from the posted token.}
\label{tab:xfam}
\footnotesize
\setlength{\tabcolsep}{5pt}
\begin{tabular}{@{}llcccc@{}}
\toprule
 & & & & At stopping point & Whole stream, $\tau=1.0$ \\
\cmidrule(lr){5-5}\cmidrule(l){6-6}
Sender & Receiver & Aligned (\%) & Flips (\%) & $\tau=0.1$ / $\tau=1.0$ / cascade (\%) & Tokens to 90\% / 99\% / 100\% \\
\midrule
Gemma 9B & Gemma 9B bf16 & 99.9 & 1.0 & 99.5 / 95.5 / 100 & 108 / 156 / 204 \\
  & Gemma 9B NF4 & 99.9 & 8.4 & 88.5 / 91.5 / 94 & 120 / 168 / 240 \\
  & Llama 8B bf16 & 88.2 & 35.5 & 1 / 19 / 19 & 336 / 528 / 636 \\
  & Llama 8B NF4 & 88.2 & 36.7 & 0.5 / 15.5 / 15.5 & 348 / 516 / 648 \\
  & Qwen 7B bf16 & 88.5 & 37.4 & 0 / 14 / 14 & 396 / 612 / 756 \\
  & Qwen 7B NF4 & 88.5 & 37.7 & 0 / 12.5 / 12.5 & 384 / 552 / 756 \\
\midrule
Llama 8B & Llama 8B bf16 & 99.9 & 1.0 & 100 / 99.5 / 100 & 96 / 120 / 180 \\
  & Llama 8B NF4 & 99.9 & 10.7 & 93 / 99 / 99 & 96 / 144 / 180 \\
  & Gemma 9B bf16 & 88.9 & 35.6 & 7.5 / 60.5 / 60.5 & 216 / 288 / 372 \\
  & Gemma 9B NF4 & 88.9 & 35.9 & 6 / 58.5 / 58.5 & 228 / 276 / 324 \\
  & Qwen 7B bf16 & 99.5 & 36.6 & 13 / 71 / 71 & 180 / 300 / 324 \\
  & Qwen 7B NF4 & 99.5 & 36.5 & 12.5 / 71 / 71 & 192 / 300 / 384 \\
\midrule
Qwen 7B & Qwen 7B bf16 & 99.9 & 1.2 & 100 / 100 / 100 & 84 / 108 / 180 \\
  & Qwen 7B NF4 & 99.9 & 10.5 & 99.5 / 100 / 100 & 96 / 120 / 180 \\
  & Gemma 9B bf16 & 91.0 & 36.0 & 12.5 / 79.5 / 79.5 & 252 / 528 / 612 \\
  & Gemma 9B NF4 & 91.0 & 36.2 & 14.5 / 81 / 81 & 240 / 516 / 588 \\
  & Llama 8B bf16 & 97.5 & 34.2 & 27.5 / 90.5 / 90.5 & 180 / 432 / 468 \\
  & Llama 8B NF4 & 97.5 & 34.9 & 27.5 / 90 / 90 & 168 / 432 / 444 \\
\bottomrule
\end{tabular}
\end{table*}
Every posted token is still the winner of the sender's race, and no acceptance test or residual sampling is needed. A bit-exact scheme could not do this unless the receiver reproduced every verification pass exactly, including which drafted tokens were accepted, since a batched pass computes the sender's probabilities along yet another path (\autoref{sec:background}). For \stealth{}, it is just one more small mismatch. We repeated the runs of \autoref{tab:main} with four drafted tokens per round, of which 26--42\% were accepted, and receivers see no difference (Appendix~\ref{app:spec}, \autoref{tab:spec}): flip rates match the plain sender's within 0.5 points for every receiver, and every same-size receiver decodes within three messages of its plain-sender count, in either direction. Smaller same-family receivers vary more, by 3--17 messages, mostly in the speculative sender's favor but not consistently (Llama 3B gains five messages on the H100 and loses five on the A6000). One Gemma message passed the sender's own check yet was missed by the same-configuration bf16 receiver: no tokens flipped, but its score sat at the edge of the beam, and a wider beam does decode it successfully.

\subsection{Cross-Family Decoding}
\label{sec:eval-xfam}

We re-ran the sweep of \autoref{sec:eval-robust} on the A6000 with each large sender's noise keyed by token bytes rather than vocabulary index (\autoref{sec:xfam}), so receivers of the other two model families can decode from the posted strings alone. \autoref{tab:xfam} gives the results, with same-family bf16 and NF4 receivers kept as a control. The byte-keyed variant costs nothing within a family: its control rows decode within one percentage point of \autoref{tab:main}'s at the sender's stopping point with the cascade, and within three points at either $\tau$ alone, as expected for an independent sample of 200 messages. They also need within two 12-token steps as much text as \autoref{tab:sweep}'s, even though \autoref{tab:xfam}'s token counts use $\tau=1.0$.

Across families, the receiver's tokenizer splits the text differently from the sender's: 88--99.5\% of its tokens line up with a sender token at the same byte offset and length, lowest whenever Gemma's tokenizer is on either side. Of the aligned tokens, 34--38\% flip, comparable to the smaller same-family receivers of \autoref{tab:main} (24--37\%), and quantizing the receiver to NF4 adds almost nothing on top, since the difference between families already dominates. Given the whole stream, every cross-family receiver decodes every message, 90\% of them within 168--252 tokens for Llama and Qwen senders and 336--396 for Gemma's.

The sender's same-model stopping check is not calibrated for these receivers: at the sender's stopping point they decode only 12.5--90.5\% of messages, almost entirely through $\tau=1.0$, since $\tau=0.1$ decodes at most 27.5\%. A sender that knows its receiver's model can instead run that model for its stopping check. Replaying the sweeps under that rule, stopping at the end of the first comment after the receiver's $\tau=0.1$ check has decoded plus 12 more tokens, the sender posts more text, and 11--88\% of messages need a second comment.

Across roughly 450,000 cross-family sweep decodes, a wrong payload was output 35 times. Each decode checks at most $B=16$ final guesses against a 16-bit CRC, which bounds false accepts at $16\cdot2^{-16}$ per decode, about 110 over these decodes; the 35 observed sit well within that bound.

\subsection{Comparison With Other Methods}
\label{sec:eval-compare}

\autoref{sec:exact} showed what mismatch does to bit-exact schemes: arithmetic coding and Discop decode every message when the receiver replays the sender's exact computation (\autoref{tab:exact}), and almost none under any mismatch, including a smaller model of the same family. \autoref{tab:compare} sets their rates against ours for context. Under replay, arithmetic coding posts 0.19--0.28 bits per token at 32-bit payloads and 0.95--0.99 at 256 bits, against our 0.19--0.26 and 0.37--0.43, and Discop is similar. Those rates hold only for a receiver that reconstructs the sender's exact distribution, which \autoref{sec:exact} found does not happen in practice.

Among receiver-free schemes, we reimplemented Zamir's~\citep{zamir2024excuse}, following its reference code, on the same news channel with a budget of 32 comments, tuning its parameters on held-out messages. It delivers 0.017, 0.024, and 0.002 payload bits per posted token on Gemma, Llama, and Qwen respectively, and decodes 92\%, 99.5\%, and 60\% of messages from its own tokens within that budget. Being receiver-free does not protect it from re-tokenization: decoding from the posted strings instead of the sender's tokens drops Llama from 99.5\% to 78.5\%, since one changed token corrupts its binary choices and every symbol's score after it. We deliver 9--15$\times$ more bits per posted token on Gemma and Llama, and nearly 100$\times$ more on Qwen, whose low-entropy text starves Zamir's per-choice evidence. We did not reimplement Bai et al.~\citep{bai2024asymmetric}. \autoref{tab:compare} instead reports their own measured rate as a rough point of comparison.

\begin{table}[t]
\centering
\caption{Payload bits per posted token. Robust: decodes despite sender--receiver mismatch. Exact: stegotext follows the sender's model distribution. --: impractical at 256 bits. $^\dagger$Rate reported by the authors.}
\label{tab:compare}
\footnotesize
\setlength{\tabcolsep}{4pt}
\begin{tabular}{@{}lcccc@{}}
\toprule
Method & Robust & Exact & 32 bits & 256 bits \\
\midrule
Arithmetic coding~\citep{ziegler2019neural} & No & Yes & 0.19--0.28 & 0.95--0.99 \\
Discop~\citep{ding2023discop} & No & Yes & 0.19--0.29 & 0.81--0.92 \\
Zamir~\citep{zamir2024excuse} & Yes & Yes & 0.002--0.024 & -- \\
Bai et al.~\citep{bai2024asymmetric}$^\dagger$ & Yes & Yes & 0.002--0.044 & -- \\
\textbf{\stealth{} (ours)} & \textbf{Yes} & \textbf{Yes} & \textbf{0.19--0.26} & \textbf{0.37--0.43} \\
\bottomrule
\end{tabular}
\end{table}

\begin{figure}[t]
\centering
\begin{tikzpicture}
\begin{axis}[
    width=0.85\columnwidth,
    height=0.5\columnwidth,
    xlabel={Payload (bits)},
    ylabel={Bits per Posted Token},
    xmode=log,
    log basis x=2,
    xmin=28, xmax=290,
    xtick={32,64,128,256},
    xticklabels={32,64,128,256},
    ymin=0, ymax=0.5,
    ytick={0,0.1,0.2,0.3,0.4,0.5},
    legend pos=south east,
    legend style={font=\footnotesize},
    label style={font=\small},
    tick label style={font=\small},
    axis lines*=left,
    axis line style={gray},
    grid=major,
    grid style={gray!20, line width=0.3pt},
    tick style={gray},
]
\addplot[line width=1.3pt, dotted, color=seriesaqua, mark=triangle*, mark size=1.4pt] coordinates {
(32,0.260) (64,0.317) (128,0.374) (256,0.408)
};
\addlegendentry{Gemma}
\addplot[line width=1.3pt, dashed, color=seriesorange, mark=square*, mark size=1.2pt] coordinates {
(32,0.223) (64,0.364) (128,0.395) (256,0.435)
};
\addlegendentry{Llama}
\addplot[line width=1.3pt, solid, color=seriesblue, mark=*, mark size=1.2pt] coordinates {
(32,0.192) (64,0.359) (128,0.352) (256,0.371)
};
\addlegendentry{Qwen}
\end{axis}
\end{tikzpicture}
\caption{Payload bits per posted token rises with payload size and then flattens, as the unused tail of the last comment shrinks relative to the whole message (median over 200 messages).}
\label{fig:capacity}
\end{figure}
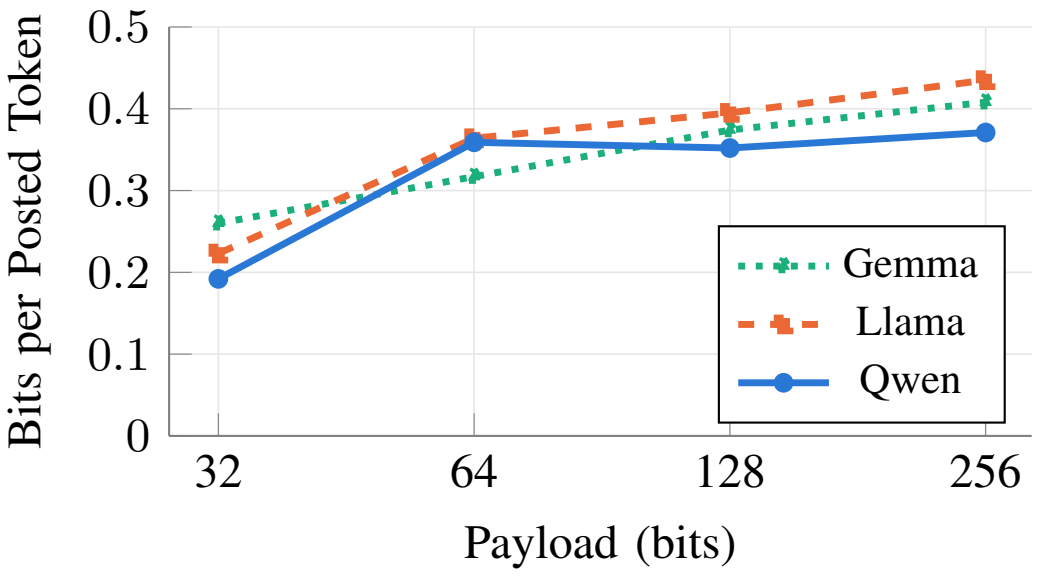

\textbf{Rate vs.\ payload size.} Our rate rises with payload size because a comment has to end where the model ends it: a short message is often decodable well before the comment it rides in is over, so most of that comment's tokens count against the rate for nothing, while a longer message uses more of the same comment before it ends. \autoref{fig:capacity} traces this for the same-model bf16 receiver from 32 to 256 bits. The rate rises by 20--90\% from 32 to 64 bits, then flattens into the 0.35--0.45 range by 128 bits, as the unused comment tail shrinks relative to a growing message instead of staying a fixed cost. It does not keep climbing toward the replay-mode rates of arithmetic coding or Discop, which spend nothing on a receiver that reproduces the sender's distribution bit for bit, while ours turns only about 40\% of the text's entropy into payload, the rest being the redundancy the beam search needs to absorb flips.

\textbf{Computational cost.} Hiding a message adds little time on either side. We timed 50 messages per sender on an A6000, one job per GPU at batch size 1. The exponential race runs within 3--4\% of the speed of honest multinomial sampling, at 25--52~ms per token for the 7--9B models, and the sender's decodability checks add another 6--7\%, for 4.2--7.0~s per message. Our receiver scores the posted text in one teacher-forced pass (44--69~ms) and runs the beam search (24--27~ms), for 70--95~ms per message on the 7--9B models. Arithmetic coding sends at the same speed but pays at the receiver: in bf16 it decodes only by regenerating the sender's text token by token, at 3.7--5.5~s per message, 50--60$\times$ ours. The float64 hardening of \autoref{sec:exact} costs more still: 2.5--5.5$\times$ our time per token on the sender, and 5.8--16.5~s per message on the receiver against our 55--71~ms on the same 1.5--3B models. Zamir's scheme also generates at the model's speed, but its low rate means a 32-bit message takes a median of 1,200--7,200 tokens, or 33--196~s to send.

\subsection{Detection Tests}
\label{sec:eval-detect}

The proof of \autoref{sec:security} covers every efficient warden; here we check that real detectors agree. Each test compares the first comment posted on each article, 2,000 per family and kind: \emph{stego}, our scheme's output; \emph{race}, an honest sample from the same exponential race run on fresh rather than keyed noise; and \emph{mult}, an honest sample from ordinary multinomial sampling, Hugging Face's default. Stego-vs-race and stego-vs-mult are the tests that matter. Race-vs-mult compares two honest samplers and shows what chance looks like. As a positive control, \emph{ac12} is arithmetic coding run deliberately at a coarse 12-bit interval precision, a scheme with known statistical artifacts.

\textbf{Trained steganalyzers.} We re-implemented three word-level linguistic steganalyzers from their published descriptions, with embeddings trained from scratch: FCN~\citep{yang2019fast}, R-BiLSTM-C~\citep{niu2019rbilstmc}, and a densely connected LSTM with feature pyramid~\citep{yang2020densely}. Alongside them we fine-tune RoBERTa~\citep{liu2019roberta} end to end as a transformer-based baseline. Each is trained and evaluated with 5-fold cross-validation grouped by article, so no article's texts appear in both training and test, and we report out-of-fold AUC with a 95\% bootstrap interval. \autoref{tab:detect-clf} gives RoBERTa, the strongest of the four on the positive control. On every stego-vs-race, stego-vs-mult, and race-vs-mult pair, RoBERTa's AUC is within 0.007 of chance and the other three are within 0.015, none significant ($|z| \le 1.3$). The control shows how little power these text-only classifiers have on distribution-preserving schemes: on only 200 texts per class, RoBERTa detects ac12 on Llama and Qwen ($z=2.93$, $2.43$) but not on Gemma, the densely connected LSTM detects it on Llama alone ($z=2.40$), and FCN and R-BiLSTM-C never do.

\begin{table}[t]
\centering
\caption{RoBERTa steganalyzer AUC (0.5 = chance). $^*$: $p<0.05$ against chance.}
\label{tab:detect-clf}
\footnotesize
\setlength{\tabcolsep}{4pt}
\begin{tabular}{@{}lccc@{}}
\toprule
Pair & Gemma & Llama & Qwen \\
\midrule
stego vs.\ mult & 0.503 & 0.498 & 0.507 \\
stego vs.\ race & 0.501 & 0.500 & 0.499 \\
race vs.\ mult (control) & 0.499 & 0.500 & 0.496 \\
ac12 vs.\ mult (positive control) & 0.515 & 0.585$^*$ & 0.570$^*$ \\
\bottomrule
\end{tabular}
\end{table}

\textbf{A detector that knows the model.} Our warden knows the sender's model (\autoref{sec:threat}), so the stronger test scores each comment with that model. For every token it computes the token's log-probability, rank, the distribution's entropy, and a randomized probability integral transform (PIT), which is uniform on $[0,1]$ under honest sampling. Each comment becomes a vector of histograms and moments of these, tail counts, PIT autocorrelation at lags 1--12 (the spine value repeats every 12 words), and length. Logistic regression, gradient boosting, and an MLP are trained on these features with the same grouped cross-validation. Across the three families and classifiers, stego-vs-mult and stego-vs-race AUCs lie in 0.48--0.52, inside the 0.48--0.54 spread of the honest race-vs-mult pair.
The same detector finds ac12 in every family, at AUC up to 0.67 ($z=6.0$).

\textbf{The missing-tail test.} \citeauthor{cao2026lpv}'s test~\citep{cao2026lpv} targets schemes that can never emit some tokens, the failure \autoref{sec:fp}'s unreachable-mass bound rules out for \stealth{}. It counts how many tokens land in the extreme tail of the sender model's distribution, below a threshold probability, and compares that count with the number honest sampling predicts, pooled into a $z$-score over a corpus. Positive $z$ is a deficit of tail tokens, the signature \citeauthor{cao2026lpv} test for; negative $z$ is an excess, which we also count as detection. \autoref{tab:cao} pools all three families, about 880,000 tokens per honest or stego kind, at three thresholds. Stego and race stay within $|z|<1.7$ at every threshold. The honest mult sampler shows a real deficit at $2^{-16}$ and $2^{-24}$ ($z=3.7$, $6.4$): \texttt{torch.multinomial} draws through a float32 cumulative sum, which under-samples the most extreme tail. The test is correctly flagging a known flaw in that particular honest sampler. Our race reads probabilities directly and has no such flaw. Arithmetic coding at 32-bit precision, the exact baseline of \autoref{sec:exact}, is borderline at $2^{-16}$ and $2^{-24}$ ($z\approx-2$). At 16 bits it is caught at $2^{-32}$ ($z=-18.5$), and at 12 bits at every threshold, down to $z=-385$.

\textbf{What these tests show.} A score near chance means only that a test with this much power found nothing, so the controls matter as much as the stego rows. Both model-aware tests catch ac12 in every family, and the missing-tail test also catches 16-bit arithmetic coding and the float32 flaw in an honest sampler. That test's power falls with its threshold. At $2^{-24}$ honest sampling expects about 74 tail tokens across the corpus, so a sampler that never reached that tail would score $z\approx 8.6$; at $2^{-32}$ it expects fewer than two, so even a fully missing tail would score only $z=1.3$, as mult does. \stealth{}'s unreachable mass, $1.4\times10^{-17}$ per token (\autoref{sec:fp}), lies far below what any such test can resolve. No finite set of tests rules out every warden; that is the job of the reduction in \autoref{sec:security}. What these results add is that detectors able to find small artifacts in other samplers find none in ours.

\begin{table}[t]
\centering
\caption{\citeauthor{cao2026lpv}'s test pooled over the three families, by tail threshold. Positive: fewer tail tokens than honest sampling predicts; negative: more.}
\label{tab:cao}
\footnotesize
\setlength{\tabcolsep}{4pt}
\begin{tabular}{@{}lccc@{}}
\toprule
Kind & $2^{-16}$ & $2^{-24}$ & $2^{-32}$ \\
\midrule
stego & 0.3 & 1.6 & -0.2 \\
race (control) & -1.1 & 1.1 & -1.6 \\
mult (control) & 3.7 & 6.4 & 1.3 \\
ac, 32-bit & -1.9 & -2.1 & 0.4 \\
ac, 16-bit & -1.0 & -1.7 & -18.5 \\
ac, 12-bit & -8.3 & -56.4 & -385.2 \\
\bottomrule
\end{tabular}
\end{table}

\section{Conclusion}
\label{sec:conclusion}

Provably secure linguistic steganography has assumed that the receiver reproduces the sender's token distribution bit for bit. \autoref{sec:exact} shows that ordinary deployments break this assumption: a different precision, quantization level, model size, or GPU leaves arithmetic coding and Discop decoding almost no messages. \stealth{} drops the assumption. Its sender picks each token by a keyed exponential race, so a receiver whose distribution differs slightly still picks the same token almost everywhere, and a disagreement costs only that one token's evidence. Because the spinal code is rateless, every further token adds evidence for the whole payload, so a receiver that disagrees more often simply needs more text.

Across three model families, this held for every receiver we tested. Given enough text, every same-family receiver, quantized or smaller and on either GPU, decodes every message, and with the noise keyed by token bytes, so does every receiver of another family. The sender is also free to generate with speculative decoding, with no measurable effect on its receivers. Robustness costs no security: \autoref{sec:security} proves the posted texts indistinguishable from honest samples of the sender's model, and no detector we tried, including one that scores text with that model, tells them apart (\autoref{sec:eval-detect}). Nor does it cost practicality: \stealth{} delivers 0.19--0.26 payload bits per posted token on 32-bit messages, an order of magnitude more than the receiver-free scheme of \citet{zamir2024excuse}, and its race runs within a few percent of honest sampling.

Sender and receiver therefore no longer need matching deployments to communicate covertly. They can run different hardware, different precisions, or different model families, and the texts they exchange remain provably indistinguishable from honest model output.

\bibliographystyle{IEEEtranN}
\bibliography{mybib}

\appendices
\section{Proofs}
\label{app:proofs}

\begin{proof}[Proof of Lemma~\ref{lem:race}]
Scaling an $\mathrm{Exp}(1)$ variable by $1/p_i(h)$ gives $E_i/p_i(h) \sim \mathrm{Exp}(p_i(h))$, so the race scores are independent exponentials with rates $p_i(h)$ summing to 1. For independent exponentials, the index achieving the minimum wins with probability its own rate over the sum of all rates, here $p_i(h)$, and the minimum itself is exponential with rate equal to that sum, here $\mathrm{Exp}(1)$, independent of the index.
\end{proof}

\begin{proof}[Proof of Lemma~\ref{lem:fresh}]
A truly random function's values at distinct inputs are independent and uniform; querying it at an input not yet read returns a value independent of every value read so far (lazy sampling). Positions strictly increase, so every noise value read before position $t$, whether to sample a token or to run a decodability check, has a first component $t' < t$: the decodability check only replays decoding over tokens already written, so it never queries position $t$ or later. No input at $t$ has therefore been queried before, so by lazy sampling its value is a fresh independent uniform, regardless of which spine value selects it.
\end{proof}

\begin{proof}[Proof of Corollary~\ref{cor:honest}]
By Lemmas~\ref{lem:race} and~\ref{lem:fresh}, each token, given everything chosen before it, is distributed exactly as $\hat p(h)$ for the current context $h$; \textsc{end} and the prompt policy are those of the honest process, so the token sequence follows the honest transition law at every step.
\end{proof}

\begin{proof}[Proof of Theorem~\ref{thm:fixed}]
We bound $\mathrm{Adv}(A)$ through a sequence of hybrids, starting at the real stego oracle and ending at the honest oracle.

$H_0$ is the real stego oracle $O_S$.

$H_1$ replaces $\mathrm{HMAC\text{-}SHA256}(K, \cdot)$ by a random function; since the warden never repeats a nonce (\autoref{def:security}), the message keys $K_m$ it induces are independent and uniform. A warden distinguishing $H_0$ from $H_1$ yields a distinguisher for HMAC-SHA256: it runs the warden, simulates the sender, and obtains each $K_m$ by querying its own oracle on the nonce $\nu$, so $|\Pr_{H_0}-\Pr_{H_1}| \le \epsilon_{\mathrm{HMAC}}(q)$.

$H_2$ replaces the ChaCha20 block function under each of the $q$ message keys by an independent random function; a standard hybrid over the $q$ messages gives $|\Pr_{H_1}-\Pr_{H_2}| \le q\,\epsilon_{\mathrm{ChaCha}}(r)$.

$H_3$ replaces the sender's implemented race by the ideal race on exact uniforms. $H_2$ and $H_3$ read the same noise words and draw the same warden coins, so they agree until the first position where the two races disagree. By Lemma~\ref{lem:fresh} the noise at that position is fresh given everything before it, so by Lemma~\ref{lem:fp} the races disagree there with probability at most $\delta$; a union bound over at most $T$ positions gives $|\Pr_{H_2}-\Pr_{H_3}| \le T\delta$. The sender's decodability checks (\autoref{sec:length}) add no term: they query only positions already sampled, and under the theorem's stopping rule their outcome does not affect what is posted.

Finally, by Corollary~\ref{cor:honest}, each message in $H_3$ is an honest conversation cut by a rule that does not use the key; the honest oracle $O_H$ applies the same rule to an honest conversation, and distinct messages use independent random functions throughout $H_1$--$H_3$, so $H_3$ and $O_H$ have exactly the same law. Chaining the four bounds gives the claim.
\end{proof}

\begin{proof}[Proof of Lemma~\ref{lem:fp}]
The strategy is to condition on a ``good'' event $\Gamma$ in which the winning margin is not too close to a tie and no noise word landed in an unusually wide cell; we bound the probability the good event fails to hold, and separately bound the implementation's disagreement probability given that it does.

Write $G_i = 1-U_i$, $X_i = E_i/\hat p_i$ with $E_i = -\ln G_i$, $i^\star = \argmin_i X_i$ and $x = X_{i^\star}$. By Lemma~\ref{lem:race}, $\Pr[i^\star=i]=\hat p_i$, $x\sim\mathrm{Exp}(1)$ is independent of $i^\star$, and given $(i^\star,x)$ the values $u_j = \hat p_j(X_j-x)$, $j\neq i^\star$, are independent $\mathrm{Exp}(1)$. Let $I_i$ be the interval of values of $-\ln(1-u)/\hat p_i$ over $u$ in $W_i$'s cell $[W_i\omega,(W_i+1)\omega)$, and $\Delta_i$ its length. Write $\hat S_i$ for the implementation's computed log score, within $\eta$ of $\ln\hat E_i - \ell_i$ by assumption. Both $X_i$ and $\hat X_i = \hat E_i/\hat p_i$ lie in $I_i$, and the computed $\tilde X_i = \exp(\hat S_i + c)$, with $c$ the normalizing constant of $\ell$, lies within a factor $e^{\pm\eta}$ of $\hat X_i$.

\emph{Cell widths.} If $G_i \ge 2\omega$, the cell is $[g,g+\omega)$ in $G$ with $g \ge \omega$, so $-\ln$ varies over it by $\ln(1+\omega/g) \le \omega/g \le 2\omega/G_i = 2\omega e^{E_i}$, and $\Delta_i \le 2\omega e^{E_i}/\hat p_i$.

\emph{Disagreement needs a close competitor.} If the implemented race picks $v \neq i^\star$,
\[
X_v - \Delta_v \le \hat X_v \le e^{\eta}\tilde X_v \le e^{\eta}\tilde X_{i^\star} \le e^{2\eta}\hat X_{i^\star} \le e^{2\eta}(x+\Delta_{i^\star}),
\]
so $X_v - x \le \Delta_v + \Delta_{i^\star} + \rho(x+\Delta_{i^\star})$.

\emph{Good event.} Let $\Gamma$ be the event that $G_i \ge 8\omega$ for all $i$, $x \le a$, and $\Delta_{i^\star} \le \tfrac12$. A union bound gives $\Pr[G_i < 8\omega \text{ for some } i] \le 8|V|\omega$, and $\Pr[x>a]=e^{-a}$. Since $E_{i^\star} = \hat p_{i^\star} x$, if $G_{i^\star} \ge 2\omega$ then $\Delta_{i^\star} \le 2\omega e^{\hat p_{i^\star} x}/\hat p_{i^\star} \le 2\omega e^x/\hat p_{i^\star}$, so $\Delta_{i^\star} > \tfrac12$ requires $\hat p_{i^\star} < 4\omega e^x$. Since $i^\star$ is independent of $x$ and $\Pr[\hat p_{i^\star} < \theta] = \sum_{i:\hat p_i<\theta}\hat p_i \le |V|\theta$, this has probability at most $4|V|\omega\,\mathbb{E}[e^x; x\le a] = 4a|V|\omega$. So $\Pr[\bar\Gamma] \le 8|V|\omega + e^{-a} + 4a|V|\omega$. It remains to bound the disagreement probability assuming $\Gamma$ holds.

\emph{On $\Gamma$.} Multiplying the disagreement condition by $\hat p_v \le 1$ and using $\hat p_v\Delta_v \le 2\omega e^{E_v} \le \tfrac14$ (as $G_v \ge 8\omega$) gives $u_v \le \tfrac12+\tfrac14+\rho(a+\tfrac12) \le 1$. Hence $E_v = \hat p_v x + u_v \le x+1$, so $\hat p_v\Delta_v \le 2e\,\omega e^x$, and
\[
u_v \le c_v := 2\omega e^x\bigl(\hat p_v/\hat p_{i^\star} + e\bigr) + \hat p_v\,\rho\,(x+\tfrac12).
\]
Given $(i^\star,x)$, $\Pr[u_j \le c_j] \le c_j$ for each $j$, and $\sum_{j\neq i^\star} c_j \le 2\omega e^x(1/\hat p_{i^\star} + e|V|) + \rho(x+\tfrac12)$. Averaging over $i^\star$ ($\mathbb{E}[1/\hat p_{i^\star}] = |\{i:\hat p_i>0\}| \le |V|$) and then over $x \le a$ ($\mathbb{E}[e^x;x\le a]=a$, $\mathbb{E}[x]+\tfrac12=\tfrac32$) bounds the disagreement probability on $\Gamma$ by $2(1+e)a|V|\omega + \tfrac32\rho$. Adding $\Pr[\bar\Gamma]$ gives the claim.
\end{proof}

\section{Speculative Decoding Results}
\label{app:spec}

\autoref{tab:spec} gives the per-receiver results behind the speculative-decoding comparison of \autoref{sec:eval-robust}. On each GPU, the sender drafts four tokens per round with a small model of its own family (Gemma 2B, Llama 1B, Qwen 0.5B) and keeps them up to the first position where its own race disagrees. Every receiver decodes the posted text at the sender's stopping point with the cascade decoder. The H100 plain columns are the runs behind \autoref{tab:main}, including the fp16 and fp32 receivers it omits, given as counts of 200 rather than percentages; the A6000 columns repeat the comparison with texts written and decoded on the A6000.

\begin{table}[h]
\centering
\caption{Plain and speculative senders on each GPU: flip rate (\%) and messages decoded of 200 at the sender's stopping point, with the cascade decoder.}
\label{tab:spec}
\footnotesize
\setlength{\tabcolsep}{2.5pt}
\begin{tabular}{@{}llcccccccc@{}}
\toprule
 & & \multicolumn{4}{c}{H100} & \multicolumn{4}{c}{A6000} \\
\cmidrule(lr){3-6}\cmidrule(lr){7-10}
 & & \multicolumn{2}{c}{Plain} & \multicolumn{2}{c}{Spec.} & \multicolumn{2}{c}{Plain} & \multicolumn{2}{c}{Spec.} \\
\cmidrule(lr){3-4}\cmidrule(lr){5-6}\cmidrule(lr){7-8}\cmidrule(lr){9-10}
Family & Receiver & Flips & Dec. & Flips & Dec. & Flips & Dec. & Flips & Dec. \\
\midrule
Gemma & 9B bf16 & 0.9 & 200 & 1.0 & 199 & 0.9 & 200 & 0.9 & 199 \\
 & 9B fp16 & 1.3 & 199 & 1.4 & 200 & 1.3 & 200 & 1.3 & 199 \\
 & 9B fp32 & 1.3 & 199 & 1.4 & 200 & 1.3 & 200 & 1.3 & 199 \\
 & 9B int8 & 3.8 & 196 & 4.0 & 199 & 4.2 & 199 & 3.9 & 198 \\
 & 9B NF4 & 8.4 & 187 & 8.9 & 188 & 8.9 & 189 & 8.7 & 186 \\
 & 2B bf16 & 31.5 & 84 & 31.2 & 101 & 31.6 & 87 & 31.4 & 90 \\
\midrule
Llama & 8B bf16 & 1.0 & 200 & 1.0 & 200 & 1.1 & 200 & 1.0 & 200 \\
 & 8B fp16 & 1.0 & 200 & 0.9 & 200 & 1.0 & 200 & 1.0 & 200 \\
 & 8B fp32 & 1.1 & 200 & 0.9 & 200 & 1.1 & 200 & 1.0 & 200 \\
 & 8B int8 & 4.9 & 200 & 4.6 & 200 & 4.7 & 200 & 4.8 & 200 \\
 & 8B NF4 & 10.2 & 196 & 10.1 & 196 & 10.3 & 197 & 10.2 & 198 \\
 & 3B bf16 & 23.5 & 175 & 23.6 & 180 & 23.6 & 185 & 23.6 & 180 \\
\midrule
Qwen & 7B bf16 & 1.2 & 200 & 1.2 & 200 & 1.2 & 200 & 1.3 & 200 \\
 & 7B fp16 & 1.3 & 200 & 1.3 & 200 & 1.4 & 200 & 1.3 & 200 \\
 & 7B fp32 & 1.3 & 200 & 1.3 & 200 & 1.3 & 200 & 1.3 & 200 \\
 & 7B int8 & 4.2 & 200 & 4.2 & 200 & 4.3 & 199 & 4.3 & 200 \\
 & 7B NF4 & 12.1 & 199 & 12.0 & 200 & 12.3 & 199 & 12.3 & 198 \\
 & 1.5B bf16 & 36.9 & 168 & 37.1 & 173 & 37.2 & 173 & 37.1 & 178 \\
\bottomrule
\end{tabular}
\end{table}




\end{document}